\documentclass[12pt,preprint]{aastex}

\shorttitle{OVI in Elliptical Galaxies}
\shortauthors{Bregman et al.}

\begin{document}

\title{OVI in Elliptical Galaxies: Indicators of Cooling Flows}

\author{Joel N. Bregman}
\affil{Department of Astronomy, University of Michigan, Ann Arbor, MI 48109}
\email{jbregman@umich.edu}
\and
\author{Eric D. Miller}
\affil{Kavli Institute for Astrophysics and Space Science, MIT, Cambridge, MA 02139}
\email{milleric@mit.edu}
\and
\author{Alex E. Athey}
\affil{The Observatories, Carnegie Institution of Washington, Pasadena, CA 91101}
\email{alex@ociw.edu}
\and
\author{Jimmy A. Irwin}
\affil{Department of Astronomy, University of Michigan, Ann Arbor, MI 48109}
\email{jairwin@umich.edu}


\begin{abstract}
Early-type galaxies often contain a hot X-ray emitting interstellar medium (3-8 $\times$10$^{6}$ K) with an
apparent radiative cooling time much less than a Hubble time.  If unopposed by a heating
mechanism, the gas will radiatively cool to temperatures $\lesssim$ 10$^{{\rm 4}}$ K at a rate proportional to $L_X$/$T_X$,
typically 0.03-1 M$_{\odot}$ yr$^{-1}$.  We can test if gas is cooling through the 3$\times$10$^{5}$ K range by observing
the OVI doublet, whose luminosity is proportional to the cooling rate.  Here we report on a study
of an unbiased sample of 24 galaxies, obtaining {\it Far Ultraviolet Spectroscopic Explorer\/} spectra to
complement the X-ray data of {\it ROSAT\/} and {\it Chandra\/}.  The OVI line emission was detected in
about 40\% of the galaxies and at a luminosity level similar to the prediction from the cooling flow
model.  There is a correlation between ${\dot{M}}_{OVI}$ and ${\dot{M}}_X$, although there is significant
dispersion about the relationship, where the OVI is brighter or dimmer than expected by a factor
of three or more.  If the cooling flow picture is to be retained, this dispersion requires that cooling
flows be time-dependent, as might occur by the activity of an AGN.  However, of detected
objects, those with the highest or lowest values of ${\dot{M}}_{OVI}$/${\dot{M}}_X$ are not systematically hot
or cool, as one might predict from AGN heating.

\end{abstract}


\section{Introduction}
The cumulative stellar mass loss from the stars in an early-type galaxy is typically 0.1-3 M$_{\odot}$
yr$^{-1}$ (e.g., \citealt{athey02}), which, integrated over a Hubble time, 
is comparable to the gaseous mass of a spiral galaxy.  The
absence of a massive disk of cool atomic and molecular gas in early-type galaxies indicates a different
life cycle for the gas, which may be divided into distinct stages.  Ignoring accretion onto a galaxy,
the first stage is mass loss from the stars, a process that can be measured by detecting the infrared
signature associated with the stellar winds of red giants.  The measurement of this process
successfully detects the infrared emission near 10 $\mu$m and yields a value for the mass loss rate that
is approximately the value predicted from theoretical stellar evolution models \citep{athey02,knapp92}.

The stellar ejecta will not remain in orbit around its star because the expanding ejecta will
eventually collide with the ejecta from other stars, undergoing shocks that convert their random
orbital motion to thermal energy (e.g., \citealt{math03}).  This process heats the gas to 
10$^{6.5}$ -10$^{7}$ K, and if there were no
additional heating or cooling, the gas would be bound to the galaxy and have the same spatial
distribution as the stars.  However, the radiative cooling time for the gas is less than a Hubble
time, so the gas will evolve with time, and this is the basis of the cooling flow model.  In the
absence of a heating mechanism, radiative cooling drains energy from the gas most rapidly at
small radii, causing a loss of buoyancy and a subsequent inflow of gas.  Then, the rate at which
gas cools and flows inward (${\dot{M}}_X$) is proportional to 
two observed quantities, the energy loss rate ($L_{X}$)
divided by the thermal energy per gram ($3k{T_{X}}/2{\mu} m_{p}$), 
or ${\dot{M}}_X$ $\propto$ $L_{X}$/$T_{X}$, typically 0.03-3 M$_{\odot}$ yr$^{-1}$  for
ellipticals (e.g., \citealt{sarz89}).  The location at which 
the gas cools (to 10$^{{\rm 4}}$ K or cooler) depends upon a free
parameter in this {\it standard\/} cooling flow picture.
This {\it standard\/} picture does not include heating effects by supernovae, which will make
profound changes, since the net cooling rate is given by 
${\dot{M}}_X$ $\propto$ ($L_{X}$-{\it H\/})/$T_{X}$, where {\it H\/} is a heating
component.  In principle, the rate at which cool gas is produced can be reduced to zero.  If one
uses the recent values for the supernova rate in early-type galaxies (\citealt{capp99}; a
factor of three lower than the older Tammann and Sandage rates; \citealt{vdbt91}), 
the characteristic temperature of the ISM would be 1$\times$10$^{7}$ K, which is about the escape
temperature for a typical elliptical galaxy ($L_{B}$ $\lesssim$ 0.3$L_{{\rm *}}$).  
More detailed hydrodynamic calculations \citep{david91,pell98,math03} show that galactic winds (or partial
galactic winds) will play an important role in the evolution of the hot gas even if the supernova
heating rate is lower than that implied by \citet{capp99}.  A galactic wind carries away
nearly all of its energy rather than radiating it, substantially lowering the X-ray luminosity.  Not
only is a galactic wind important for understanding the X-ray emission and the ISM in the galaxy,
it has implications for the surrounding intergalactic medium, as this is the primary way that it
becomes polluted with metals.

The X-ray emission surveys of early-type galaxies show that there can be a wide range in
the mass and luminosity of X-ray emitting gas for galaxies of similar optical properties (e.g.,
\citealt{brown98,osull01}).  This variation, along with the
trend of rapidly decreasing $L_{X}$ with decreasing optical luminosity $L_{B}$, can be understood if galactic
winds play a role in some systems (other systems may have accretion from their surrounding
region; review by \citealt{math03}).  Therefore, the prediction is that the systems
with large X-ray gas masses and short cooling times (high $L_{X}$) should have cooling flows while
the X-ray poor systems have partial or total galactic winds, so the cooling flow will be weak or
absent in those systems.

Because the heating rate is not well known, it is difficult to determine the net cooling rate
through X-ray luminosity measurements.  Alternatively, if gas is cooling to lower temperatures,
emission lines will be produced that are indicative of those lower temperatures.  The best lines for
this test are from OVI because this ionization state dominates the total radiative cooling as the gas
passes through the 2-4$\times$10$^{5}$ K range \citep{edgar86}.  The primary cooling lines
come from the doublet at $\lambda$$\lambda$1032, 1038 \AA , which is accessible with the {\it Far Ultraviolet
Spectroscopic Explorer\/} ({\it FUSE\/}).  Calculations show that there is a linear relationship between the
line luminosity and the cooling rate, which is insensitive to the metallicity of the gas or whether
the gas is out of collisional ionization equilibrium \citep{edgar86,voit94}.  
At 10$^{5.3}$ K, the radiative cooling rate per unit volume is significantly higher than at
the temperature of the X-ray emitting gas (10$^{6.5}$-10$^{7}$ K), so the relative contribution of a heating
mechanism is greatly reduced and can be ignored.  Therefore, the luminosity of the OVI doublet
should be a direct measure of the gas cooling rate, ${\dot{M}}$.

Previously, we reported upon OVI observations of two early-type galaxies, NGC 1404
and NGC 4636 \citep{breg01}, two of the X-ray luminous
galaxies widely believed to host cooling flows.  For NGC 1404, OVI was not detected and the
upper limit is several times below the standard cooling flow prediction, based on {\it ROSAT\/} data. 
However, OVI emission was detected from NGC 4636, and the luminosity of these lines
corresponds to a cooling rate of 0.4 M$_{\odot}$ yr$^{-1}$.  This is less than the total 
rate from the cooling flow model of 2 M$_{\odot}$ yr$^{-1}$, but 
the {\it FUSE\/} aperture (a 30$\arcsec$ square aperture, or an effective radius of about
17$\arcsec$) only takes in a part of the galaxy.  Correcting for the flux that falls outside the aperture is a
model-dependent procedure, but if one uses a model with distributed mass drop out ({\it q\/}=1 
from \citealt{sarz89}), the corrected OVI luminosity approximately equals the cooling flow
prediction.

Following on this work, we began a OVI emission line survey of an unbiased sample of
nearby early-type galaxies.  The basic observations define the emission line characteristics of the
sample, and permit us to test a few predictions of the model.  One would expect that the X-ray
faint galaxies would be very weak OVI emitters, if most of the thermal energy is being carried
away in galactic winds.  Secondly, the galaxies with significant hot gas masses should usually
possess OVI emission.

\section{Sample Selection}
We defined a sample of optically selected early-type galaxies in order to be unbiased with
respect to X-ray properties.  In previous work with {\it ROSAT\/} observations \citep{brown98,brown00}, 
we developed a complete optically selected sample of early-type galaxies, which
contains the optically brightest elliptical galaxies (by total apparent magnitude) in the \citet{faber89}
sample that do not have AGNs pointing at us, are not at low Galactic latitude, and
excluding dwarf galaxies and cD systems.  This sample of 33 galaxies have been studied
extensively in the optical region and were observed at X-ray energies with {\it ROSAT\/}.  Most of the
objects in the sample were observed more recently with {\it Chandra\/}.  This sample was accepted as a
program on {\it FUSE\/}, and nearly all were observed, with only one exception due to the pointing
constraints following the reaction wheel failure.  The redshift range of the sample is -250 km s$^{-1}$ to
1950 km s$^{-1}$, and the apparent magnitude range is 9.3-11.0 (in B); for completeness, we included
M87.  The optical properties and exposure times are given in Table 1, where we list 
the galaxy name, absolute blue magnitude, R$_e$, one-dimensional stellar 
velocity dispersion, extinction in B, distance (in km s$^{-1}$ and Mpc, 
total exposure time, and night exposure time (boldface denotes the data used).

The raw {\it FUSE\/} data were processed using the pipeline CALFUSE 2.2.1, 2.4, and 3.0, usually with
very little change between the data sets.  We
inspected each of the data sets to determine whether using the night-only data was superior to
using the total data set (night plus day data).  The decisions depended partly on the redshift of the
galaxy since the weaker OVI line at $\lambda$1037.62 can be redshifted into any of three airglow lines at
1039.2 \AA , 1040.8 \AA , and 1041.6 \AA \ at about 450 km s$^{-1}$, 920 km s$^{-1}$, and 1150 km s$^{-1}$.  These
airglow lines are about 100 km s$^{-1}$ wide and can shift relative to the source depending upon the
placement in the slit.  The ratio of these airglow lines, and relative to the stronger one at 1027.3,
allows us to estimate the degree to which contamination occurs.  Also, there is a dead spot in the
detector that covers part of the spectrum near 1043 \AA \ and this sometimes produces artifacts in the
spectrum (false absorption or emission features), so we inspect this wavelength region closely in
the event that an emission line is found nearby (the weaker OVI lines occurs here for redshifts of
1500-1600 km s$^{-1}$).

There are four detectors that cover this wavelength region, two LiF channels and two SiC
channels, but the effective area is significantly better for the LiF channels.  Also, the LiF1a
channel is significantly better than the LiF2b channel and we find that adding this second channel
generally reduces the S/N, so we usually rely upon the LiF1a data for the region covering the OVI
lines.  We inspect the other channels for confirmation of features.  The other important emission
line in the {\it FUSE\/} band is from CIII $\lambda$977.02, which is covered by the two SiC channels, with the
SiC2a being the best.  Sometimes there is an offset in the wavelength scale, and when this is an
issue, we mention it and use the strong atomic absorption line, CII $\lambda$1036.34 to realign the
spectrum.

\section{Data Processing and Observational Results} 
The spectra from early-type galaxies have several components that need to be modeled or
recognized when trying to determine OVI line strengths.  Given the redshift range of our objects,
the OVI lines fall in the region 1030-1045 \AA , but these lines are superimposed upon the
continuum of the host galaxy, plus there is absorption by Milky Way gas, and there are airglow
lines (discussed above), all of which are taken into account when analyzing line strengths.

The underlying stellar spectrum of the host galaxy was modeled in this spectral region by
\citet{brown97} in order to analyze {\it Hopkins Ultraviolet Telescope\/} ({\it HUT\/}) data, which have a
resolution of about 3 \AA , significantly worse than {\it FUSE\/}.  When we compare their model spectrum
to the stellar continuum from NGC 1399, the galaxy with the brightest continuum, we find that
the model reproduces the general properties of the spectrum.  However, there are aspects of the
spectrum that depart from the model, and since the S/N is good for NGC 1399, we choose to use
it as the template spectrum for other objects.

There are important absorption lines, produced by the Milky Way, superimposed upon the
observed spectra of these external galaxies.  The strongest atomic line in the spectral range 1028-1053 \AA\  is CII $\lambda$1036.34, which is always optically thick.  This is followed in line strength by OI
$\lambda$1039.23, ArI $\lambda$1048.22, and the Galactic OVI lines.  In addition to atomic absorption lines, there
can be absorption by molecular lines, although the strength of this component can vary greatly,
from being absent (at our typical S/N) to being optically thick.  The lines occur in bands, one
occurring in the 1036-1039 \AA \ region and another that occurs in the 1048.5-1052 \AA  \ region.  The
three strongest lines in the group at 1036-1039 \AA \ are centered at 1036.54 \AA , 1037.15 \AA , and
1038.16 \AA , the first of which can blend with the atomic line of CII $\lambda$1036.34.  These atomic and
molecular lines can absorb part or all of an OVI emission line from the external galaxy if the
redshift coincidence is unfortunate.  We show the location of these absorption components in the
figures for the spectra, but we do not determine their absorption equivalent widths, as this
generally does not bear upon the analysis.

To extract line fluxes or upper limits, we defined the continuum for about 5 \AA \ on either
side of the line, along the shape of the stellar continuum; over this narrow range, the continuum is
nearly a simple slanted line.  To determine the line flux, we integrated over the line, but to
determine the FWHM, a Gaussian was fit.  The location of the continuum is one of the main
contributors to the uncertainty of the line flux, so the continuum location was varied ($\pm$ 1$\sigma$) to
determine the errors in the line flux and line width.  When the line flux was less than 3$\sigma$, we
consider it a non-detection and quote an upper limit of 3$\sigma$.

The individual objects are discussed, beginning with NGC 1399, which was used as the
stellar template.  Extracted fluxes or upper limits to the OVI lines are 
given in Table 2, and for the detections, possible detections, and most upper limits, the spectra 
are shown (some non-detections are largely noise so we do not show their spectra).  For uniformity, and to avoid the
strong geocoronal Ly$\beta$ airglow line, we show the 1028-1053 \AA \ region in our figures of all
objects.  The spectra were binned by five pixels to a bin size of 0.034 \AA \ (9.8 km s$^{-1}$ at the OVI
lines) and then smoothed by five, seven, or 11 for the purposes of producing the figures shown
below; line fluxes were extracted from the unsmoothed data. 
The reddenings quoted below are from \citet{schleg98}, the Galacitc HI column density 
is from \citet{dickey90}, and other galaxy properties (e.g., redshift) are taken from the
{\it NASA/IPAC Extragalactic Data Base\/}. 

NGC 1399: This galaxy has the brightest stellar continuum, typically an order of magnitude
brighter than other systems.  As with other galaxies in the Fornax cluster (NGC 1316, NGC
1404), the Galactic HI column is low (1.31$\times$10$^{20}$ cm$^{-2}$) as is the extinction,
and there is no evidence for Galactic H$_{2}$ absorption, although Galactic atomic absorption is found (Figure 1).  The lack of H$_{2}$ absorption simplifies the determination of the location of the
continuum and we see no evidence for OVI emission.  Due to the absence of OVI emission and
the strength of the continuum, we use the stellar continuum as the standard template for the other
systems, after filling in the Galactic absorption lines (for guidance, we used the model by \citealt{brown97}).

NGC 1316: This is also known as Fornax A, a radio galaxy in the Fornax cluster, and one of the
optically most luminous galaxies in the cluster.  It is also a LINER with weak low ionization
emission lines, extended ionized and neutral gas, and some evidence of dust.  The system has a
low Galactic extinction and low Galactic HI column (1.9$\times$10$^{20}$ cm$^{-2}$), so the
absence of detectable Galactic H$_{2}$ is expected.  The {\it FUSE\/} Lif1a spectrum (Figure 2) shows
emission from the strong line of OVI ($\lambda$1032).  There is weaker emission from the OVI ($\lambda$1038)
line, also seen in the Lif2b channel.  The sharp feature that is coincident with Ly$\beta$ for NGC 1316
and the OVI ($\lambda$1032) line from the Milky Way is not confirmed in the Lif2b channel and may not
be real (also, the higher series Ly lines are not seen).  In addition, there is emission from the CIII
($\lambda$977) line at the redshift of NGC 1316 (Sic2a and Sic1a channels).

The two OVI lines and the CIII line have the same FWHM of about 1.5 \AA \ (440 km sec$^{-1}$),
which is comparable to the FWHM that one would infer from the one-dimensional velocity
dispersion of the stars in NGC 1316 (223 km s$^{-1}$, which would lead to a FWHM of 525 km s$^{-1}$).
The line ratio of the two OVI lines is 2.7$\pm$1.0, consistent with the value of 2 expected for optically
thin gas.

NGC 1395: Galactic absorption is present but there is little H$_{2}$ absorption (about 10$^{17.5}$ cm$^{-2}$) and
there is no evidence for Galactic OVI emission, which would be difficult to detect in this short
observation.  The strong OVI line from NGC 1395 would lie at 1037.84 \AA , a fairly clean part of
the spectrum; there is only an upper limit to this line (Figure 3).

NGC 1404: Like the other Fornax galaxies, H$_{2}$ absorption is nearly absent, and in this
observation, using night data, there airglow lines are nearly absent as well (Figure 4); the usual Galactic
atomic absorption lines are apparent (presented and discussed in \citealt{breg01}).  
There is no emission from either OVI line at the redshift of NGC 1404.

NGC 1407: There is a moderate Galactic HI column (5.42$\times$10$^{20}$ cm$^{-2}$) and extinction, 
and the Galactic H$_{2}$ lines are prominent, along with the atomic absorption lines.  The
stronger OVI line from NGC 1407 (1038.1 \AA ) corresponds to the location of a Galactic H$_{2}$ line,
although not one of the strongest ones, but there is no apparent emission, nor is there emission at
the location of the weaker OVI line (Figure 5).

NGC 1549: There is very little Galactic absorption or extinction (1.57$\times$10$^{20}$ cm$^{-2}$, 
and there is little evidence of H$_{2}$ absorption (Figure 6).  Absorption by Galactic atomic gas
is present and the CII $\lambda$1036 line (0.6 \AA \ wide) lies in the middle of the redshifted OVI $\lambda$1032 line,
although emission near this region is seen in the Lif2b channel as well.  The weaker line, OVI
$\lambda$1038, is in an uncontaminated region and it is detected with a FWHM of about 1.2 \AA .  The flux
is elevated just to the blue side of the CII $\lambda$1036 line, which may be the unabsorbed part of the
OVI $\lambda$1032 line.  The line width is about 0.7 of the FWHM of the stellar velocity dispersion. There is no emission from CIII $\lambda$977.  There is a narrow emission line at the location of the
Galactic OVI line.

NGC 3115: Galactic molecular absorption lines are strong, but the redshifted OVI lines lie in
uncontaminated parts of the spectrum.  There is no emission from OVI (Figure 7).

NGC 3379: The redshift of this galaxy places the OVI lines in a part of the spectrum
uncontaminated by Galactic absorption.  No OVI emission is found (Figure 8).

NGC 3585: This fairly isolated E7/S0 galaxy lies in a loose group and has moderate Galactic reddening and
an HI column of 5.6$\times$10$^{20}$ cm$^{-2}$. There were problems that occurred when this
spectrum was being obtained, and of the 16 ksec exposure, only 3 ksec was acceptable, nearly all
during the day (Figure 9).  Consequently, the airglow lines are strong.  Absorption by Galactic atomic and
molecular gas is nearly absent, which is surprising given the column density of the gas.  Due to the
absence of the strong Galactic C II, Ar I and Fe II (1063.18 \AA ) lines, we do not believe this to be
a reliable spectrum.  This object may be reobserved at a future date but this data only provides a
poorly constrained upper limit.

NGC 3607: The Galactic HI column and extinction are low toward this system (1.52$\times$10$^{20}$ cm$^{-2}$) and there are no strong H$_{2}$ absorption lines that can be identified.  The only
clear Galactic atomic absorption line is the CII $\lambda$1036 feature, and the strong OVI line would be
redshifted to 1035.2 \AA , where there appear to be an emission feature (Figure 10).  This feature also
occurs in the lower S/N Lif2b channel, which does not add much to the S/N of this result, but
provides some consistency.  However, the emission from the weaker OVI line is not present,
although since this line is only half the strength of the other OVI line, it's absence does not lead to
an inconsistency with the presence of the other line.  The center of the OVI line is at 1035.3 \AA , a
redshift of 979 km sec$^{-1}$ (and a FWHM of 0.5$\pm$0.1 \AA , or 150 km sec$^{-1}$), which is similar to the
redshift of the galaxy, 935 km sec$^{-1}$.  In addition to the weak OVI line, the CIII $\lambda$977 line is also
detected at a somewhat higher level of significance and at exactly the redshift of the galaxy.  It has
a width of 0.72$\pm$0.15 \AA , which is consistent with the width of the OVI line.  A line width of 0.6 \AA \ 
corresponds to 180 km sec$^{-1}$, which is narrower than the FWHM that would be inferred from the
stellar velocity dispersion (248 km sec$^{-1}$, or a FWHM of 584 km sec$^{-1}$).

NGC 3923: This galaxy has significant extinction and a moderately large HI column 
(6.3$\times$10$^{20}$ cm$^{-2}$), although the Galactic H$_{2}$ lines are not especially strong.  No emission lines
are seen (Figure 11).  The weaker OVI line is in an uncontaminated part of the spectrum, while the
stronger OVI line would coincide with a weak Galactic H$_{2}$ absorption line and is adjacent to
strong airglow lines.

NGC 4125: The stellar continuum is very weak in this galaxy, despite one of the longest
observations in the sample (Figure 12).  The continuum is so poorly defined that no Galactic absorption lines
can be identified.  The night-only data (40\% of the total exposure time) set yields stricter upper
limits to the OVI emission.  No emission features from any lines are detected, despite this galaxy
being listed as a LINER and classified as E6 pec.

NGC 4374:  This elliptical in the Virgo cluster has radio lobes and this is classified as a LINER. 
Despite being at high Galactic latitude ({\it b\/} = 74$^{\circ}$), it has moderate absorption and
the {\it FUSE\/} spectrum (Figure 13) shows clear evidence for Galactic H$_{2}$ absorption.  The day
spectrum had particularly strong airglow lines, so we used the night-only spectrum, which had
only 3.7 ksec, yet it reveals a strong OVI $\lambda$1032 at the galaxy redshift (also present at the same
level in the day+night spectrum); the red side of the line may be partly absorbed by the Galactic
CII $\lambda$1036 line.  The weaker OVI line lies between two airglow lines and there is an instrumental
feature near 1044 \AA , but a line half of the strength of the OVI $\lambda$1032 line is consistent with the
level of the spectrum above that expected from the stars.  The line width of 130 km sec$^{-1}$
(FWHM) is only about one-fifth the velocity dispersion of the stars, an equivalent FWHM of 674
km sec$^{-1}$, although we cannot rule out that the line has a broader base.

NGC 4406: Also known as M86, this lies in the central part of the 
 cluster, 1.2$^{\circ }$ west of
M87, and with a modest HI column (2.6$\times$10$^{20}$ cm$^{-2}$).  It has a nearly radial orbit,
as its redshift (-244 km s$^{-1}$) differs from the systemic velocity of the cluster by more than the
cluster velocity dispersion.  It has an elongated X-ray distribution that has been interpreted as
stripping of the gas within NGC 4406 by the ambient cluster medium.  The stellar continuum is
very faint so that it is difficult to detect any Galactic absorption features (Figure 14).  The
stronger OVI line would be shifted to 1031.1 \AA , and a possible emission feature in this low S/N
spectrum occurs at that location. Although this is the broadest and most significant emission
feature, it is less than a 3$\sigma$ detection.  The weaker OVI line is not detected, nor are any other
emission lines.

NGC 4472: Known as M49, this lies in the center of the southern grouping of the Virgo cluster. 
It has a modest Galactic HI column and extinction (1.65$\times$10$^{20}$ cm$^{-2}$), but it has
clear Galactic H$_{2}$ absorption.  There is an indication of a OVI $\lambda$1032 line associated with NGC
4472, although if present, it is redshifted relative to the systemic velocity of the galaxy by about
0.4 \AA \ (120 km sec$^{-1}$; Figure 15).  There is a minor peak at the location of the redshifted OVI
$\lambda$1038 line in this night-only spectrum.  We regard the OVI line emission as a possible detection. 
The line width of FWHM = 120 km sec$^{-1}$ is about 18\% of the stellar velocity dispersion expressed
as a FWHM, 676 km sec$^{-1}$.  There is no indication of emission from the CII $\lambda$977 line.

NGC 4486: This central galaxy of the Virgo cluster (M87) has low extinction but measurable lines
of Galactic atomic and molecular gas.  Unfortunately, the Galactic CII $\lambda$1036 line (1036.3 \AA ) is
coincident with the redshifted OVI $\lambda$1032 line (at 1036.4 \AA ) and there are Galactic H$_{2}$ lines to the
red side of the Galactic CII line, so much of a redshifted OVI $\lambda$1032 line would be absorbed
(Figure 16).  Nevertheless, to the blue side of the Galactic CII line, the continuum rises well
above the expected stellar continuum (in both the Lif1a and Lif2b spectra), suggestive of the blue
side of a wide OVI $\lambda$1032 line from M87.  If this is the case, the line width would needs to be
about 1.5 \AA \ (FWHM) and at least 1.2 \AA .  If it is 1.5 \AA , we estimate a line strength of 7$\times$10$^{-15}$ erg
cm$^{-2}$ sec$^{-1}$, which is consistent with the upper limit to the weaker OVI line of 5$\times$10$^{-15}$ erg cm$^{-2}$
sec$^{-1}$.

The CIII $\lambda$977 line is detected, with a line width of 1.4 \AA , although it could be wider since
the red side is absorbed by Galactic H$_{{\rm 2}}$ lines; ignoring absorption by these lines, the line flux is
1.5$\times$10$^{-14}$ erg cm$^{-2}$ sec$^{-1}$, stronger than the OVI line.  If the H$_{2}$ absorption produces the decrease in the red side of the line, the line flux and line width would be about 30\% larger.

NGC 4494: The redshifted OVI $\lambda$1032 line (1036.6 \AA ) lies in the Galactic CII absorption line
(1036.34 \AA ), but there is no emission line on either side of the CII line (Figure 17).  There is no
evidence for the redshifted OVI $\lambda$1038 line either, although it falls on an instrumental feature in
the Lif1a channel.  In the less sensitive Lif2b channel, there is no indication of either OVI line, and
there is no instrumental feature near the weaker line.

NGC 4552: This galaxy (M89) also lies in the Virgo cluster, but has slightly more extinction than
most other members (2.56$\times$10$^{20}$ cm$^{-2}$) and shows clear evidence of Galactic
molecular hydrogen absorption.  Although Galactic atomic and molecular absorption
compromises our ability to detect the redshifted OVI $\lambda$1038 line, the redshifted OVI $\lambda$1032 line is
in a part of the spectrum that is uncontaminated by Galactic absorption or airglow (Figure 18). 
There is an emission feature near the redshift of the systemic velocity of the galaxy (for the OVI
$\lambda$1032 line), although the center of the emission feature is redshifted relative to the systemic
velocity by about 0.5 \AA \ (150 km sec$^{-1}$; the stellar velocity dispersion is 260 km sec$^{-1}$, or an
equivalent FWHM of 615 km sec$^{-1}$, so the line shift is well within the velocity dispersion value). 
The OVI $\lambda$1032 line is visible on both the Lif1a and Lif2b channels.

NGC 4621: This E5 galaxy (M59) lies about 3$^{\circ }$ west of M87 in the Virgo cluster, it has about the
average extinction for the Virgo cluster (2.2$\times$10$^{20}$ cm$^{-2}$), but it clearly shows
Galactic H$_{2}$ absorption along with the usual Galactic atomic absorption lines.  However, the
redshift of the galaxy places both OVI lines in uncontaminated regions and the spectrum is of
moderately good S/N (Figure 19).  There is no evidence for emission lines of either OVI line or of
CIII.

NGC 4636: This Virgo cluster galaxy was presented by \citet{breg01}, who
reported the detection of both OVI lines.  The width of the emission lines but not the line strength
seems to depend upon the version of the pipeline used; these results employ pipeline version 2.4 (Figure 20). 
The CIII line may be detected, although the centroid is 120 km sec$^{-1}$ (0.4 \AA )  blueward of the
expected line center (the line would have a FWHM of 0.5 \AA \ and a flux of 5$\times$10$^{-15}$ erg cm$^{-2}$ sec$^{-1}$,
detected at about the 2$\sigma$ level).

NGC 4649: The redshifted OVI $\lambda$1032 line would be shifted (1035.8 \AA ) just to the blue side of
the Galactic CII $\lambda$1036 line (1036.3 \AA ), but there is no evidence for OVI emission above the level
of the rather strong stellar continuum, which is similar to the continuum of NGC 1399 (Figure
21).  There are moderately strong Galactic H$_{2}$ lines and atomic absorption lines present, although
these do not contaminate the locations of the OVI lines.

NGC 4697: The redshift of this galaxy would place the redshifted OVI $\lambda$1032 line (1036.2 \AA ) into
the strong Galactic CII absorption line (1036.34 \AA ).  There is no strong line to the blue side of the
CII line (Figure 22).

NGC 5846: This is the optically most luminous galaxy in the center of its group and it has a
moderate amount of extinction (4.24$\times$10$^{20}$ cm$^{-2}$), 
with evidence of absorption from Galactic molecular hydrogen and atomic lines.  
This is one of the more distant objects, so
the continuum is fairly faint but is well-fit by the stellar continuum of NGC 1399 (Figure 23). 
Both the strong and weak redshifted OVI lines appear to be present, but not in the expected 2:1
ratio (for OVI $\lambda$1032 to OVI $\lambda$1038).  The OVI $\lambda$1032 line is broken up by Galactic molecular
absorption features, so it could be significantly stronger.  We cannot fit the H$_{2}$ parameters well
enough to make an accurate absorption correction for the OVI $\lambda$1032 line.  The usually weaker
OVI $\lambda$1038 line (redshifted to 1043.6 \AA ) falls close to a feature in the detector.  Whereas this
feature is stationary and well-known, we have seen instances of unusual absorption or emission
near it, and this may contribute to the strength of this feature.  The spectra in the Lif2b channel,
while poorer, confirms the presence of the OVI $\lambda$1032 line but not the OVI $\lambda$1038 line. 
Consequently, we estimate the OVI luminosity from the OVI $\lambda$1032 line.  There is emission from
the CIII $\lambda$977 line at a flux of 7.7$\times$10$^{-14}$ erg cm$^{-2}$ sec$^{-1}$ (15\% error), with a line width of 1.0$\pm$0.1 \AA \ 
and a line center of 982.87 \AA \ (at the systemic redshift, it would have been at 982.61 \AA ).

IC 1459: This galaxy has among the lowest Galactic HI column in the sample and very little
reddening (1.19$\times$10$^{20}$ cm$^{-2}$), although it has the usual Galactic atomic absorption
lines that can be seen on the stellar continuum (Figure 24).  The airglow lines are strong during
the daytime observing, so although only 25\% of the observing time was at night (only 2.5 ksec),
the night-time data are better for the analysis of the OVI region, while the total data set is better
for analysis in the CIII region.  There is a broad emission line coincident with the strong redshifted
OVI line, which is relatively uncontaminated by Galactic absorption features (also present in the
Lif2b channel).  The weaker OVI line is not detected at a statistically significant level, but as it is
half the strength of the OVI $\lambda$1032 line, this is not inconsistent with a detection of the OVI $\lambda$1032
line.  The CIII $\lambda$977 line is also detected with a flux that is poorly determined because it is unclear
where to mark the boundaries of the line.  If we use the single peak, to the blue of line center, the
line width is 0.8 \AA \ (line flux of 1.3$\times$10$^{-14}$ erg cm$^{-2}$ sec$^{-1}$) and a line center at 982.0 \AA \ (expected to
be at 982.5 \AA ).  If we use the weaker red part of the line as well, the line center is exactly that
expected for the recession velocity, the line width becomes 2.1 \AA \ (at a line flux of 2.1$\times$10$^{-14}$ erg
cm$^{-2}$ sec$^{-1}$).  Given the uncertainties associated with each of the lines, it is difficult to compare the
OVI and CIII lines.  This galaxy has a strong radio source and is known to be a LINER.

\section{Analysis of the Sample}
An important goal of the program is to test whether the X-ray cooling rate is a good
predictor of the true cooling rate, which would be reflected in the OVI line luminosity, presumed
to be a better measure of the net cooling rate.  There certainly is not a one-to-one relationship
between the X-ray value of ${\dot{M}}$ (${\dot{M}}_X$) and the OVI value of ${\dot{M}}$ (${\dot{M}}_{OVI}$), such as for the galaxies NGC 1399
and NGC 1404, two of the most X-ray luminous sources but with no detectable OVI emission. 
However, there are statistical correlations between the X-ray and OVI cooling rates.

For the analysis of our sample, we have summarized the results in Table 3, where we give
a variety of X-ray and optical properties, along with the OVI results.  The measure of a detection
is given by either upper limits, detections, or possible detections.  The detections are lines whose
strength exceeds 3$\sigma$ and where the signal appears clearly on the spectrum.  Possible detections are
generally near 3$\sigma$ threshold or where it has been difficult to establish the presence of the line
unambiguously.  There are a variety of issues, such as contamination from Galactic absorption
line, airglow features, or uncertainties in the level of the underlying stellar continuum.  
For the purposes of scoring the detections in a simple non-parametic fashion, we assign 0 to upper limits,
0.5 to possible detections, and 1.0 to detections.

\subsection{Determination of the OVI and X-Ray Cooling Rates\/}

The conversion of the OVI luminosity to a cooling rate (${\dot{M}}_{OVI}$) has been calculated
by \citet{edgar86} and by \cite{voit94}, and here we use the
conversion L(1032) = 9$\times$10$^{38}$ ${\dot{M}}_{OVI}$ erg/sec, where ${\dot{M}}_{OVI}$ is in M$_{\odot}$ yr$^{-1}$ (as used in
BMI).  This is probably accurate to 30\%, where the uncertainty derives from issues such as
whether the cooling gas is isobaric or isochoric.  This is most likely an uncertainty in the absolute
calibration and is unlikely to cause an additional random scatter of 30\%.  The quantity
${\dot{M}}_{OVI}$ is fairly insensitive to the metallicity provided that the metals remain the primary
cooling agent (Edgar and Chevalier 1986).  In the temperature range where OVI is the most
prominent ion (near 10$^{5.3}$ K), this requirement is satisfied for metallicities greater than 10$^{-2}$ of the
Solar value (see \citealt{suth93}).

We wish to compare the OVI cooling rate to the X-ray cooling rate within the {\it FUSE\/}
aperture, and this depends somewhat on whether gas cools primarily in the center or in a
distributed manner throughout the galaxy.  Effectively, the X-ray cooling rate is 
${\dot{M}}_{X} = \eta L_{\rm net}/E$, where $L_{\rm net}$ is the net cooling rate 
(the X-ray luminosity, if there is no opposing heat
source) and E is the specific thermal energy of the gas.  The factor $\eta$ represents the model
correction due to fluid effects, such as gravitational reheating as gas flows inward or a sink of
mass as gas cools out of the flow, which can be extracted from model calculations, such as
\citet{sarz89}.   We discussed the differences between the predictions in the various
models previously (BMI) and here we use the value $\eta$ = 0.4, corresponding to the {\it q \/}= 1 model.
In this {\it q \/}= 1 model, gas loss is distributed through the galaxy in the sense that hot gas is 
converted locally to cold gas at a rate inversely proportional to the instantaneous
cooling time $t_c$ as given by ${\dot{\rho}} = q{\rho}/t_c$.

There is a second consideration in calculating ${\dot{M}}_X$, having to do with the {\it FUSE\/}
aperture, a square 30$\arcsec$ on a side, or an effective radius of 17$\arcsec$.  This is smaller than the effective optical
radius (R$_{e}$, typically 30-60$\arcsec$) or of the extent of the X-ray emission, which is similar to the optical
size.  If we were to adopt the cooling flow model without mass drop out, all of the cooling would
occur within the {\it FUSE\/} aperture (although this makes an X-ray surface brightness profile that is
too sharply peaked in the center).  If there is mass drop-out with radius, then we need to correct
for only enclosing part of the cooling gas in the {\it FUSE\/} aperture.  As described in BMI, this is
accomplished by multiplying the total ${\dot{M}}_X$ by the fraction of the X-ray luminosity projected
into the {\it FUSE\/} aperture.  Here we use the value of ${\dot{M}}_X$ for the distributed model ({\it q=1\/}),
corrected to the size of the {\it FUSE\/} aperture.  If we had used the model with no mass drop-out ({\it q=0\/}), the
median predicted value of ${\dot{M}}_X$ hardly changes, although for the most X-ray luminous galaxies
in the sample (often the largest), the predicted ${\dot{M}}_X$ could be a factor of two higher.  The
various values of ${\dot{M}}$ and other relevant derived quantities, including the 
X-ray luminosity, $L_X$, and the X-ray temperature, $T_X$ are given in Table 3.

\subsection{A Correlation Between the OVI and X-Ray Cooling Rates\/}

There is a complete set of {\it ROSAT\/} X-ray fluxes and luminosities for this sample, so we
begin by comparing these to the OVI data.  The nominal prediction was that there would be a
connection between ${\dot{M}}_X$ and the detection of OVI emission, but we examined other
relationships as well.  Some of those relationships investigated were between OVI and $L_X$, $L_X/T_X$,
and $T_X$, where no strong correlations were found and no strong relationships were predicted. 
However, there seems to be a correlation between the OVI and ${\dot{M}}_X$, as seen in Figure
25.  We see that none of the six galaxies with the lowest values of ${\dot{M}}_X$ have any OVI emission, yet the significance of this correlation is difficult to quantify.  We would like to use
the Kaplan-Meier estimator for the analysis of censored data \citep{isobe86},
but an underlying assumption is that the data are censored randomly.  Here, the upper limits
(censored data) are not randomly distributed, but preferentially occur at the lower values of
${\dot{M}}_X$.  Also, we have introduced ``possible'' detections, which is difficult to incorporate in
statistical schemes.  Given these challenges, we can divide the sample into three bins by ${\dot{M}}_X$,
using our scoring for detections, possible detections, and upper limits.  We find that for galaxies
with the lowest ${\dot{M}}_X$ values, 1/8 have OVI, while 4.5/8 of the highest ${\dot{M}}_X$ objects have
OVI (and 3.5/8 of the intermediate objects have OVI).  Using Poisson statistics, the joint
probability that of the lowest ${\dot{M}}_X$ objects, 1/8 (or fewer) have OVI while 4.5/8 (or more) of
the highest ${\dot{M}}_X$ objects have OVI would occur by chance 2.5\% of the time (97.5\%
confidence level).  The correlation is probably a bit stronger than this value since none of the
lowest six ${\dot{M}}_X$ object have either an OVI detection or possible detection.  For a Kendall's $\tau$
test or a Spearman's $\rho$ test for the whole sample (treating upper limits and detections equally), the
significance improves to 98-99.5\% confidence, with the higher significance if the very poor upper
limit object, NGC 3585 is eliminated from the sample.  Conservatively, we conclude that the
correlation exists at the 98\% confidence level when using the {\it ROSAT\/} data.

A correlation is suggested between the OVI detectability and the radio luminosity, where
the lowest third (in $L_{\rm radio}$) has 1.5/8 OVI detections, the middle third has 3/8 and the most
luminous third has 4.5/8 OVI detections.  When we divide the sample in halves instead of thirds,
for the low radio luminosity group, only 1.5/12 have OVI detections while 7.5/12 of the radio
luminous objects have OVI emission.  This correlation is significant at the 93\% confidence level
(using Poisson statistics), so more objects would be required to confirm this result.  If the
correlation is real, it could be caused if the cooling gas eventually feeds the radio source, although
there is not a good correlation between ${\dot{M}}_X$ and $L_{\rm radio}$.  Alternatively, it might be possible
for the AGN to heat the gas and produce OVI emission, and four of the radio luminous objects
have CIII emission as well, although this could be consistent with cooling gas also.

The {\it Chandra\/} data offer a number of advantages in determining X-ray properties, relative
to the {\it ROSAT\/} data.  First, the point sources can be removed from the observations, and a
correction can be made for the unresolved point source contribution, yielding a more accurate
measure of the properties of the X-ray emitting gas \citep{athey03,athey05}.  
The most important observational quantities used to calculate ${\dot{M}}_X$ are $L_X$ and $T_X$
(${\dot{M}}_X \propto L_X/T_X$), with the surface brightness distribution, usually described by a $\beta$ model,
being of less importance.  The {\it Chandra\/} data not only provide better values for $L_X$, but also for
$T_X$ and for the $\beta$ model as well.  However, the values of $T_X$ are typically in the range of 0.3-0.8 keV, so it is important to have observations that cover this low energy range adequately, and
this is provided by the ACIS-S chip, which has sensitivity to 0.3 keV.  The other detector used for
imaging, ACIS-I, is three times less sensitive in the 0.5-1 keV range, it is insensitive below 0.5
keV, and it was subject to radiation damage during the early part of the mission. For these
reasons, we do not include ACIS-I data in this program for galaxies where {\it ROSAT\/} data indicated
that $T_X$ $<$ 0.5 keV.  This affects two objects, NGC 1395 and NGC 3607, which are excluded
from this analysis, making 22 objects with useful {\it Chandra\/} and {\it FUSE\/} data.  This sample includes
ACIS-I data for NGC 4486 (M87), a hotter galaxy, and we used the values given in the work of
\citet{mats02}.

Using the {\it Chandra\/} data, the relationship between ${\dot{M}}_{OVI}$ and ${\dot{M}}_X$ is little changed 
(Figure 26).  The objects with the lowest values of ${\dot{M}}_X$ are smaller
in the {\it Chandra\/} data due to the removal of the X-ray point sources.  The significance of the
correlation is a bit lower for the {\it Chandra\/} data, mainly because of having two fewer objects (95\%
confidence of a correlation).

\section{Interpretation of the Correlation and its Properties\/}

The nominal cooling flow prediction is that there should be a relationship between
${\dot{M}}_{OVI}$ and ${\dot{M}}_X$, but the relationship should be ${\dot{M}}_{OVI}$ $\approx$ ${\dot{M}}_X$.  This is nearly the
case for the {\it ROSAT\/} data and is almost the case for the {\it Chandra\/} data, save ${\dot{M}}_X$ appears to be
about 20-30\% too large, which is within the uncertainties in the model predictions.  The more
important discrepancy with the models is the very large scatter of ${\dot{M}}_{OVI}$ for a fixed ${\dot{M}}_X$
(the dispersion of the relationship).  This occurs because there are several X-ray bright galaxies
without OVI emission, such as NGC 4649 (the upper limit on ${\dot{M}}_{OVI}$ is 5 times less than the
prediction), NGC 1399 (2.6 time less than the prediction) and NGC 3923 (2 times below the
prediction).  These differences are much greater than the statistical uncertainties in the data.  Also,
there are some galaxies in which OVI is surprisingly bright: IC 1459 lies an order of magnitude
above the predicted ${\dot{M}}_{OVI}$, NCG 1549 is five times above the prediction; and NGC 1316 is
four times above the predicted value.  Even if one were to suggest that an AGN were responsible
for the high values of OVI, we are still left with the objects that are much fainter than predicted.

This broad dispersion about the relationship could be caused if cooling flows varied in
time, as has been suggested for the cluster case (e.g., \citealt{kais03}).  Heating by an
AGN could reheat the gas, choking off the cooling flow.  If that were the case, one might expect
that, in the moderate and high ${\dot{M}}_X$ objects without OVI emission, the gas temperature would
be above the velocity dispersion temperature, indicating that net heating is occurring.  Similarly,
the OVI emitters, at similar values of ${\dot{M}}_X$, should have temperatures at or below the stellar
velocity dispersion temperature.  However, we find no such trend in the data supporting this
picture nor is there any correlation between the detection of OVI and $T_{X}$ $\sigma$$^{-2}$ for the entire sample. 
Also, one might expect a division by radio properties, assuming that this is a proxy for AGN
activity.  Yet the objects furthest above the line, ${\dot{M}}_{OVI}$ = ${\dot{M}}_X$, are not obviously
different in their radio properties than those furthest below that line.  Therefore, within the
context of the cooling flow picture, time variation in the flow rate is necessary to explain the OVI
detection properties, yet we do not find telltale signs that should accompany intermittent heating.

Another possibility is that the OVI emission is not a result of cooling gas, but a result of
gas being heated, such as by thermal conduction.  However, there is a very large difference in the
emission line strength between gas that is radiatively cooling and gas that is being conductively
heated.  This difference, which is typically a factor of 10$^{3}$-10$^{4}$ (Canizares et al. 1993), means that
to achieve the same line flux for conductively heated gas, the mass flux rate would have to be
typically 10$^{2}$-10$^{3}$ M$_{\odot}$ yr$^{-1}$ instead of a cooling rate of 0.1-1 M$_{\odot}$ yr$^{-1}$.  If this cold gaseous
reservoir is HI or H$_{2}$, the mass would have to be less than typical upper limits for the galaxies in
this sample, about 10$^{8.5}$ M$_{\odot}$, with some galaxies 
having significantly stricter upper limits \citep{roberts91,breg92}.  At a conduction rate of 10$^{2.5}$
M$_{\odot}$ yr$^{-1}$, this would lead to a lifetime for the cold gas reservoir of 10$^{6}$ yr, so the gas would have
to be replenished every 10$^{6}$ yr by the amount of gas found in galaxies about one-tenth the mass of
the Milky Way.  The orbital interaction time would be about 10$^{8}$ yr, so we should see many of
these galaxies passing through the early-type galaxy, which is not the case.  Consequently, it
seems unlikely that conductive heating of gas could produce the lines that we observe.

A final issue is whether the gas is undergoing simple radiative cooling or whether turbulent
mixing layers play a role \citep{slavin93}.  In this case, the mixing would be
between the hot ambient medium and gas that has already cooled, and this process can cause gas
to spend less time at a given temperature compared to the pure radiative model.  This has been
suggested as a process within cluster cooling flows in order to remain consistent with the
discrepancy between the observed and predicted strength of the X-ray OVII line within the
cooling flow model (Fabian et al. 2001).  A similar discrepancy exists for NGC 4636, where the
OVII line is weaker than expected for gas cooling at the rate derived from the {\it FUSE\/} data (Xu et
al. 2003).  If turbulent mixing causes gas to effectively jump over the OVII temperature region
($\sim$ 10$^{6}$ K), it could solve this issue, but it also makes the prediction that there will be emission from
the CIV $\lambda$ 1550 line at a luminosity greater than the OVI doublet.  Unfortunately, there is
presently no instrument that can measure the  CIV $\lambda$ 1550 line with the required sensitivity.

\section{Future Prospects\/}

The use of OVI emission has given a new insight into the properties of the hot gas in
early-type galaxies, and this study highlights the need for further investigations.  The greatest need
is for a substantial increase in sensitivity, since most of the galaxies were not detected and even
the detections are typically at the 3-5$\sigma$ level.  For objects with short exposure times ($<$ 5 ksec), it
should be possible to double or even triple the S/N with moderate investments of observing time
(this applies to six objects in the sample, four of which are detections or possible detections).  The
other objects generally have exposure times of 5-10 ksec, requiring an additional 15-30 ksec per
object (generally of night data) to double the S/N.  For the objects with upper limits, improving
the S/N by 2-3 will not make these objects secure detections even if there is a weak feature
present.  Improved studies of these objects, or of more distant sources will require a new
instrument with at least an order of magnitude greater sensitivity.  This instrumental goal should
be achievable as {\it FUSE\/} is a rather small instrument by the standards of any space-based optical
telescope (or even most X-ray telescopes).

\acknowledgements
We would like to thank the {\it FUSE\/} team for their assistance in the collection and reduction
of these data, and in particular, B-G Andersson for his patience in dealing with many data
reduction issues.  Also, we wish to acknowledge the advice of Renato Dupke and Bill Mathews. 
This research has made use of the NASA/IPAC Extragalactic Database (operated by JPL, Caltech),
the High Energy Astrophysics Science Archive, the Multimission Archive at Space Telescope, and
the NASA Astrophysics Data System, all operated under contract with NASA.
We gratefully acknowledge support by NASA through grants NAG5-9021, NAG5-11483, G01-2089, GO1-2087, GO2-3114, and NAG5-10765.

\begin{deluxetable}{rrrrrrrrrr}
\tablecolumns{10}
\tablewidth{0pc}
\tablecaption{Observational Properties of Galaxy Sample}
\tablehead{
\colhead{Galaxy} & \colhead{Vel} & \colhead{B$_{0T}$} & \colhead{R$_e$} & \colhead{$\sigma $} 
& \colhead{A(B)} & \colhead{D} & \colhead{D} & \colhead{t$_{exp}$(tot)} & \colhead{t$_{exp}$(Nt)} \\ 
\colhead{} & \colhead{(km/s)} & \colhead{(mag)} & \colhead{($\arcsec$)} & \colhead{(km s$^{-1}$)} & 
\colhead{(mag)} & \colhead{(km s$^{-1}$)} & \colhead{(Mpc)} & \colhead{(sec)} & \colhead{(sec)} }
\startdata
NGC 1316 & 1760 & 11.34 & 80.1 & 174 & 0.09 & 1422 & 20.3 & 18861 & \bf{9166} \\ 
NGC 1395 & 1717 & 11.02 & 45.1 & 258 & 0.10 & 1990 & 28.4 & 16573 & \bf{9875} \\ 
NGC 1399 & 1425 & 10.55 & 42.4 & 310 & 0.06 & 1422 & 20.3 & 16523 & \bf{10956} \\ 
NGC 1404 & 1947 & 10.89 & 26.7 & 225 & 0.05 & 1422 & 20.3 & \bf{7751} & 6363 \\ 
NGC 1407 & 1779 & 10.57 & 72.0 & 285 & 0.30 & 1990 & 28.4 & 16282 & \bf{10287} \\ 
NGC 1549 & 1220 & 10.58 & 47.4 & 205 & 0.05 & 1213 & 17.3 & 17900 & \bf{11868} \\ 
NGC 3115 & 720 & 9.95 & 32.3 & 266 & 0.21 & 1021 & 14.6 & 11561 & \bf{8080} \\ 
NGC 3379 & 911 & 10.43 & 35.2 & 201 & 0.11 & 857 & 12.2 & 4002 & \bf{2565} \\ 
NGC 3585 & 1399 & 10.53 & 38.0 & 220 & 0.28 & 1177 & 16.8 & \textbf{2998} & \bf{2998} \\ 
NGC 3607 & 935 & 10.53 & 65.5 & 248 & 0.09 & 1991 & 28.4 & 12918 & \bf{9131} \\ 
NGC 3923 & 1788 & 10.52 & 52.2 & 216 & 0.36 & 1583 & 22.6 & \bf{15667} & 10420 \\ 
NGC 4125 & 1356 & 10.57 & 58.4 & 229 & 0.08 & 1986 & 28.4 & 25129 & \bf{9366} \\ 
NGC 4374 & 1060 & 10.13 & 54.5 & 287 & 0.17 & 1333 & 19.0 & 6444 & \bf{3692} \\ 
NGC 4406 & -244 & 9.87 & 101.7 & 250 & 0.12 & 1333 & 19.0 & 5378 & \bf{4248} \\ 
NGC 4472 & 997 & 9.32 & 103.6 & 287 & 0.10 & 1333 & 19.0 & 5381 & \bf{3732} \\ 
NGC 4486 & 1307 & 9.52 & 103.6 & 361 & 0.10 & 1333 & 19.0 & 3612 & \bf{3612} \\ 
NGC 4494 & 1351 & 10.69 & 45.1 & 124 & 0.09 & 695 & 9.9 & 18616 & \bf{12679} \\ 
NGC 4552 & 340 & 10.84 & 30.0 & 261 & 0.18 & 1333 & 19.0 & \bf{11013} & 8623 \\ 
NGC 4621 & 410 & 10.65 & 45.5 & 240 & 0.10 & 1333 & 19.0 & \bf{6747} & 3413 \\ 
NGC 4636 & 938 & 10.20 & 101.1 & 191 & 0.12 & 1333 & 19.0 & 6457 & \bf{5137} \\ 
NGC 4649 & 1117 & 9.77 & 73.4 & 341 & 0.11 & 1333 & 19.0 & 15771 & \bf{7192} \\ 
NGC 4697 & 1241 & 10.03 & 73.5 & 165 & 0.13 & 794 & 11.3 & 11712 & \bf{9446} \\ 
NGC 5846 & 1714 & 10.67 & 82.6 & 278 & 0.24 & 2336 & 33.4 & 9636 & \bf{7739} \\ 
IC 1459 & 1691 & 10.88 & 34.0 & 308 & 0.07 & 2225 & 31.8 & 10358 & \bf{2563} \\ 
\enddata
\end{deluxetable}


\begin{deluxetable}{rrrrrrrrr}
\tablecolumns{9}
\tablewidth{0pc}
\tablecaption{OVI and CIII Measurements}
\tablehead{
\colhead{Galaxy} & \colhead{OVI 1032} & \colhead{OVI 1038} & \colhead{Detect} & \colhead{F(OVI 1032)} & \colhead{Err} & \colhead{FWHM} & \colhead{Err} & \colhead{CIII} \\
\colhead{} & \colhead{(\AA )} & \colhead{(\AA )} &  & \colhead{erg s$^{-1} $ cm$^{-2}$} & \colhead{erg s$^{-1} $ cm$^{-2}$} & \colhead{(\AA )} & \colhead{(\AA )} &  \\ }
\startdata
NGC 1316 & 1037.99 & 1043.71 & Y & 1.1E-14 & 2.0E-15 & 1.5 & 0.2 & Y \\ 
NGC 1395 & 1037.84 & 1043.56 & UL & 1.5E-15 &  &  &  & N \\ 
NGC 1399 & 1036.84 & 1042.55 & UL & 3.0E-15 &  &  &  & N \\ 
NGC 1404 & 1038.63 & 1044.36 & UL & 3.0E-15 &  &  &  & N \\ 
NGC 1407 & 1038.05 & 1043.78 & UL & 7.0E-16 &  &  &  & N \\ 
NGC 1549 & 1036.13 & 1041.84 & Y & 4.6E-15 & 8.0E-16 & 1.2 & 0.2 & N \\ 
NGC 3115 & 1034.41 & 1040.11 & UL & 1.5E-15 &  &  &  & N \\ 
NGC 3379 & 1035.07 & 1040.77 & UL & 2.5E-15 &  &  &  & N \\ 
NGC 3585 & 1036.75 & 1042.46 & UL & 1.0E-14 &  &  &  & N \\ 
NGC 3607 & 1035.15 & 1040.86 & Y & 1.6E-15 & 7.0E-16 & 0.5 & 0.1 & Y \\ 
NGC 3923 & 1038.08 & 1043.81 & UL & 1.0E-15 &  &  &  & N \\ 
NGC 4125 & 1036.60 & 1042.31 & UL & 8.0E-16 &  &  &  & N \\ 
NGC 4374 & 1035.58 & 1041.29 & Y & 4.0E-15 & 6.0E-16 &  &  & N \\ 
NGC 4406 & 1031.09 & 1036.78 & P & 3.8E-15 & 1.5E-15 & 1 & 0.2 & N \\ 
NGC 4472 & 1035.36 & 1041.07 & P & 3.2E-15 & 1.0E-15 & 0.4 & 0.1 & N \\ 
NGC 4486 & 1036.43 & 1042.14 & P & 7.0E-15 & 3.0E-15 &  &  & P \\ 
NGC 4494 & 1036.58 & 1042.30 & UL & 1.0E-15 &  &  &  & N \\ 
NGC 4552 & 1033.10 & 1038.80 & P & 3.0E-15 & 1.0E-15 & 1 & 0.2 & N \\ 
NGC 4621 & 1033.34 & 1039.04 & UL & 3.0E-15 &  &  &  & N \\ 
NGC 4636 & 1035.16 & 1040.87 & Y & 3.9E-15 & 6.0E-16 & 1 & 0.2 & P \\ 
NGC 4649 & 1035.77 & 1041.49 & UL & 9.0E-16 &  &  &  & N \\ 
NGC 4697 & 1036.20 & 1041.92 & UL & 1.4E-15 &  &  &  & N \\ 
NGC 5846 & 1037.83 & 1043.55 & Y & 3.0E-15 & 1.0E-15 & 0.6 & 0.2 & Y \\ 
IC 1459 & 1037.75 & 1043.47 & Y & 9.5E-15 & 4.0E-15 & 1.5 & 0.5 & Y \\ 
\enddata
\tablecomments{Y signifies a detection, P a possible detection, and UL an upper limit.}
\end{deluxetable}

\begin{deluxetable}{rrrrrrrrrrr}
\rotate
\tablecolumns{11}
\tablewidth{0pc}
\tablecaption{Cooling Rates}
\tablehead{
\colhead{} & \colhead{} & \colhead{FUSE} & \colhead{FUSE} & \colhead{ROSAT} & \colhead{ROSAT} & \colhead{ROSAT} & \colhead{Chandra} & \colhead{Chandra} & \colhead{Chandra} &  \\ 
\colhead{Galaxy} & \colhead{OVI} & \colhead{logL(OVI)} & \colhead{${\dot{M}}_{OVI}$} & \colhead{log$L_X$} & \colhead{$T_{X}$} & \colhead{${\dot{M}}_X$} & \colhead{$L_{X,gas}$} & \colhead{$T_{X,gas}$} & \colhead{${\dot{M}}_X$} & \colhead{F(1.4 GHz)} \\ 
\colhead{} & \colhead{} & \colhead{erg s$^{-1}$} & \colhead{M$_{\odot}$ yr$^{-1}$} & \colhead{erg s$^{-1}$} & \colhead{keV} & \colhead{M$_{\odot}$ yr$^{-1}$} & \colhead{erg s$^{-1}$} & \colhead{keV} & \colhead{M$_{\odot}$ yr$^{-1}$} & \colhead{Jy}
}
\startdata
NGC 1316 & Y & 38.85 & 0.78 & 40.79 & 0.35 & 0.15 & 40.77 & 0.58 & 0.20 & 120 \\ 
NGC 1395 & UL & 38.29 & 0.21 & 40.75 & 0.44 & 0.18 &  &  &  & UL \\ 
NGC 1399 & UL & 38.24 & 0.19 & 41.15 & 0.94 & 0.22 & 41.16 & 0.77 & 0.48 & 2.5 \\ 
NGC 1404 & UL & 38.23 & 0.19 & 40.98 & 0.56 & 0.32 & 40.54 & 0.63 & 0.13 & 0.0039 \\ 
NGC 1407 & UL & 38.20 & 0.18 & 41.05 & 0.91 & 0.12 & 41.00 & 0.68 & 0.22 & 0.088 \\ 
NGC 1549 & Y & 38.29 & 0.21 & 39.75 & 0.19 & 0.040 & 39.70 & 0.29 & 0.041 & UL \\ 
NGC 3115 & UL & 37.84 & 0.08 & 39.45 & 0.45 & 0.011 & 38.88 & 0.61 & 0.004 & 0.0012 \\ 
NGC 3379 & UL & 37.78 & 0.07 & 39.49 & 0.26 & 0.019 & 39.15 & 0.61 & 0.006 & 0.0024 \\ 
NGC 3585 & UL & 38.87 & 0.83 & 39.55 & 0.31 & 0.018 & 39.59 & 0.53 & 0.020 & UL \\ 
NGC 3607 & Y & 38.30 & 0.22 & 40.53 & 0.37 & 0.095 &  &  &  & 0.0069 \\ 
NGC 3923 & UL & 38.23 & 0.19 & 40.61 & 0.55 & 0.092 & 40.63 & 0.29 & 0.37 & UL \\ 
NGC 4125 & UL & 37.99 & 0.11 & 40.72 & 0.33 & 0.18 & 40.46 & 0.37 & 0.11 & 0.0025 \\ 
NGC 4374 & Y & 38.46 & 0.32 & 40.80 & 0.70 & 0.11 & 40.72 & 0.50 & 0.16 & 2.95 \\ 
NGC 4406 & P & 38.37 & 0.26 & 41.72 & 0.67 & 0.53 & 41.35 & 0.67 & 0.24 & UL \\ 
NGC 4472 & P & 38.26 & 0.20 & 41.48 & 0.94 & 0.21 & 41.26 & 0.72 & 0.28 & 0.22 \\ 
NGC 4486 & P & 38.60 & 0.44 & 42.00 & 0.80 & 0.84 & 42.40 & 1.70 & 0.25 & 38.5 \\ 
NGC 4494 & UL & 37.19 & 0.02 & 38.99 & 0.30 & 0.005 & 38.78 & 0.48 & 0.004 & 0.0035 \\ 
NGC 4552 & P & 38.33 & 0.24 & 40.63 & 0.43 & 0.18 & 40.39 & 0.55 & 0.13 & 0.1 \\ 
NGC 4621 & UL & 38.24 & 0.19 & 39.50 & 0.63 & 0.007 & 39.25 & 0.63 & 0.007 & UL \\ 
NGC 4636 & Y & 38.38 & 0.27 & 41.52 & 0.72 & 0.31 & 41.26 & 0.68 & 0.24 & 0.079 \\ 
NGC 4649 & UL & 37.73 & 0.06 & 41.19 & 0.82 & 0.18 & 41.09 & 0.76 & 0.32 & 0.029 \\ 
NGC 4697 & UL & 37.50 & 0.03 & 39.84 & 0.30 & 0.007 & 39.30 & 0.60 & 0.003 & UL \\ 
NGC 5846 & Y & 38.90 & 0.87 & 41.72 & 0.73 & 0.61 & 41.55 & 0.63 & 0.53 & 0.021 \\ 
IC 1459 & Y & 39.14 & 1.55 & 40.90 & 0.6 & 0.22 & 40.30 & 0.59 & 0.11 & 0.97 \\ 
\enddata
\tablecomments{Y signifies a detection, P a possible detection, and UL an upper limit.}
\end{deluxetable}

\clearpage

\begin{figure}
\plotone{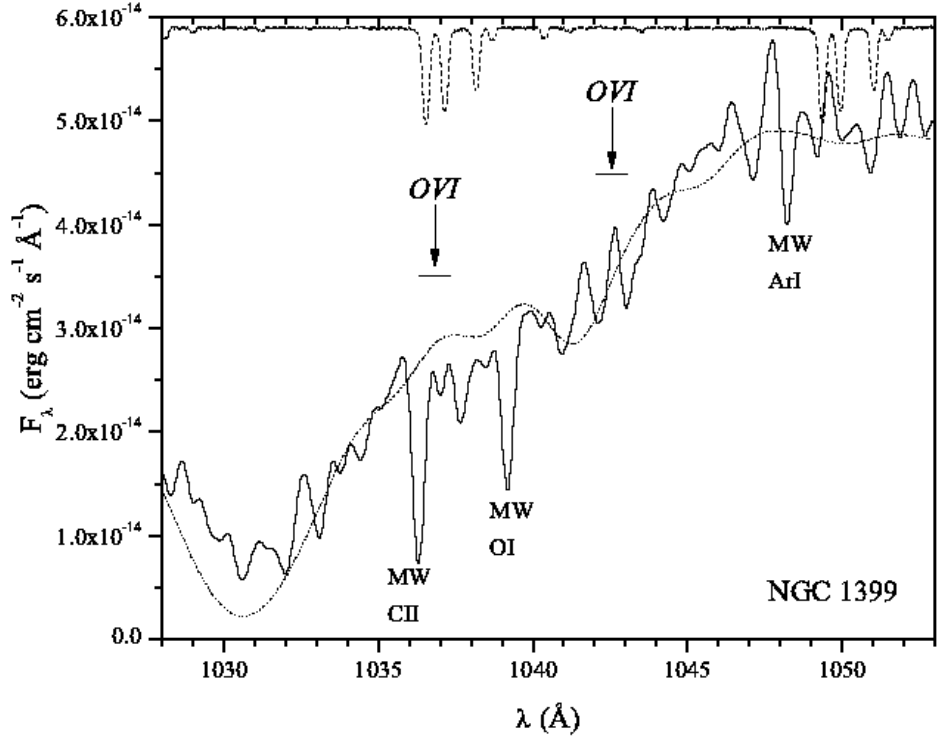}
\caption{The {\it FUSE\/} spectrum of NGC 1399, using the Lif1a channel is 
the solid line.  The dotted line is a stellar population model by \citet{brown97}.
The dashed line at the top shows where Galactic H$_2$ absoption would occur (none
detected) and the strong Galactic atomic absorption lines are labeled at the bottom
; MW denotes Milky Way features.
The location of the redshifted OVI lines are marked and emission lines are not 
detected.\label{fig1}}
\end{figure}

\begin{figure}
\plotone{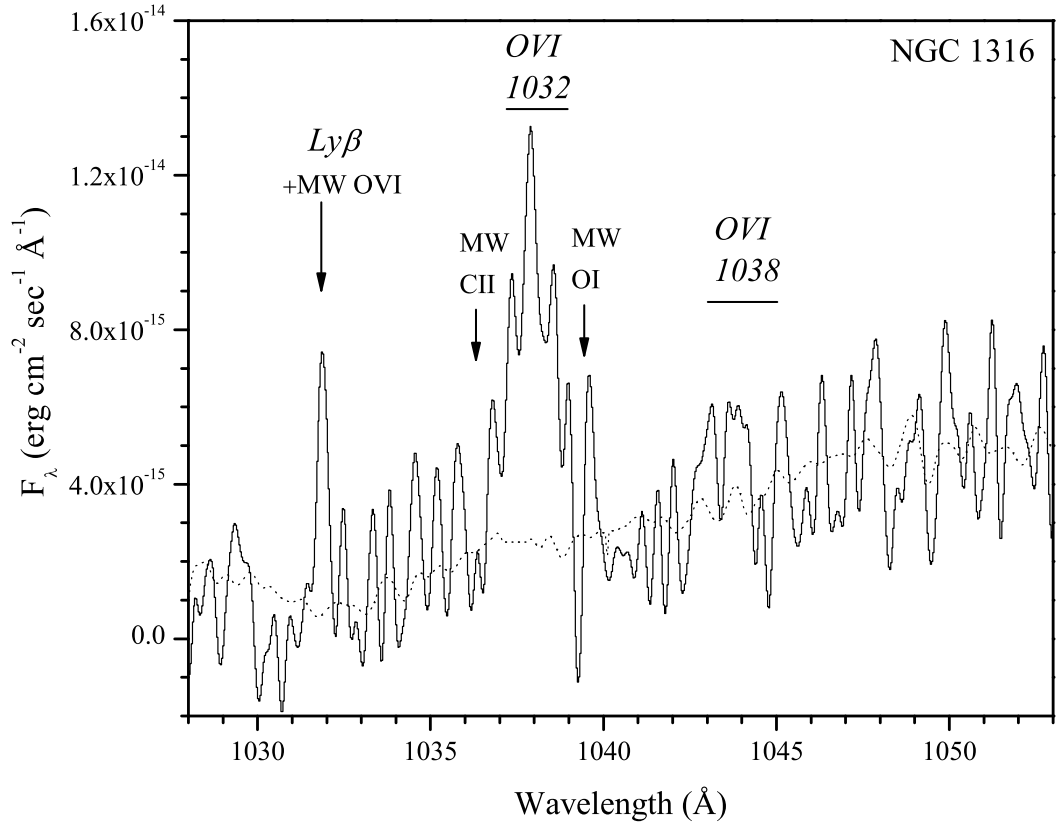}
\caption{The spectrum of NGC 1316 shows strong OVI emission (labeled in italics)
and a weak stellar continuum (the dashed line is the stellar continuum of NGC 1399).  
Galactic atomic absorption is seen but not H$_2$ absorption at this S/N.\label{fig2}}
\end{figure}

\begin{figure}
\plotone{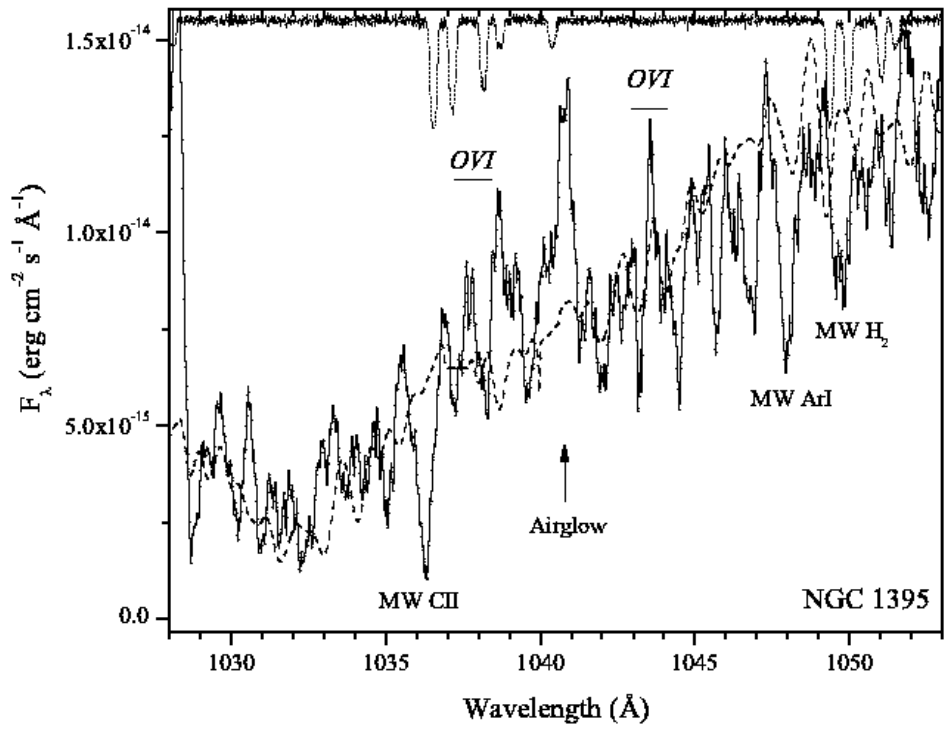}
\caption{The spectrum of NGC 1395 does not show OVI emission at a statistically
significant level.  Galactic atomic absorption is apparent along with 
weak H$_2$ absorption.\label{fig3}}
\end{figure}

\begin{figure}
\plotone{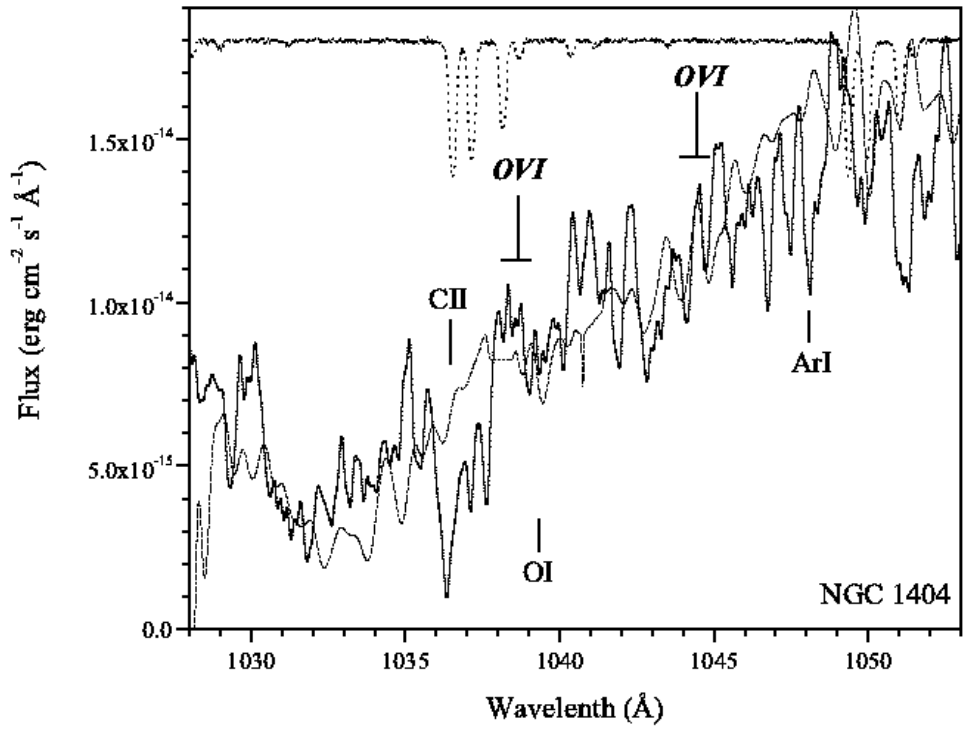}
\caption{The spectrum of NGC 1404 is well-fit by the stellar continuum of NGC 1399,
plus Galactic absorption lines, but without OVI emission at a statistically
significant level.\label{fig4}}
\end{figure}

\begin{figure}
\plotone{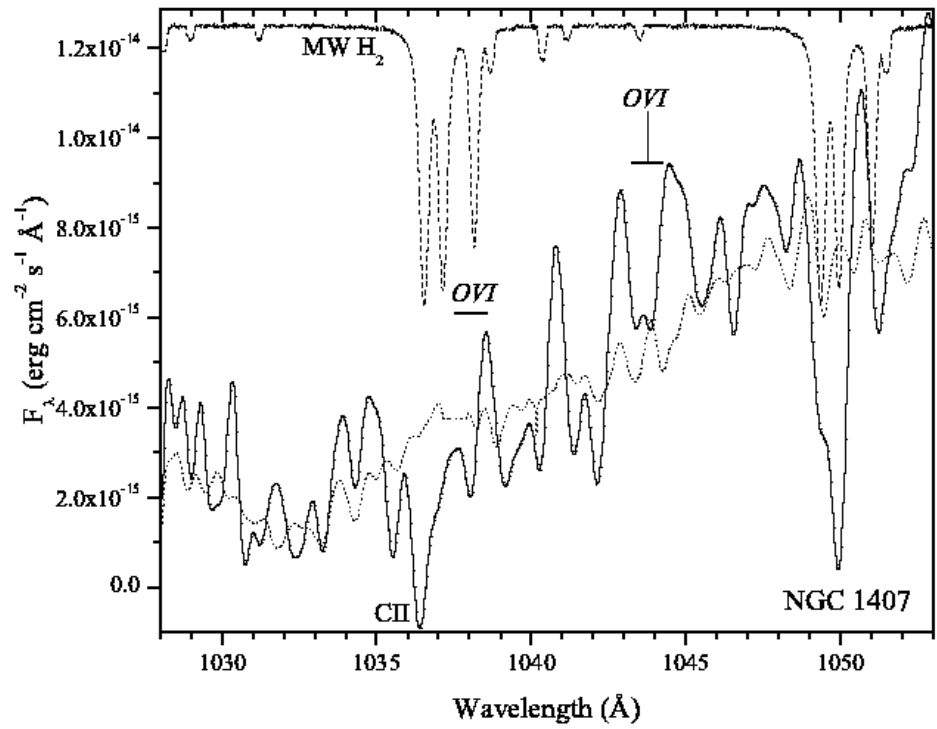}
\caption{The spectrum of NGC 1407 does not show OVI emission at a statistically
significant level.  Galactic atomic and molecular absorption is detected.\label{fig5}}
\end{figure}

\begin{figure}
\plotone{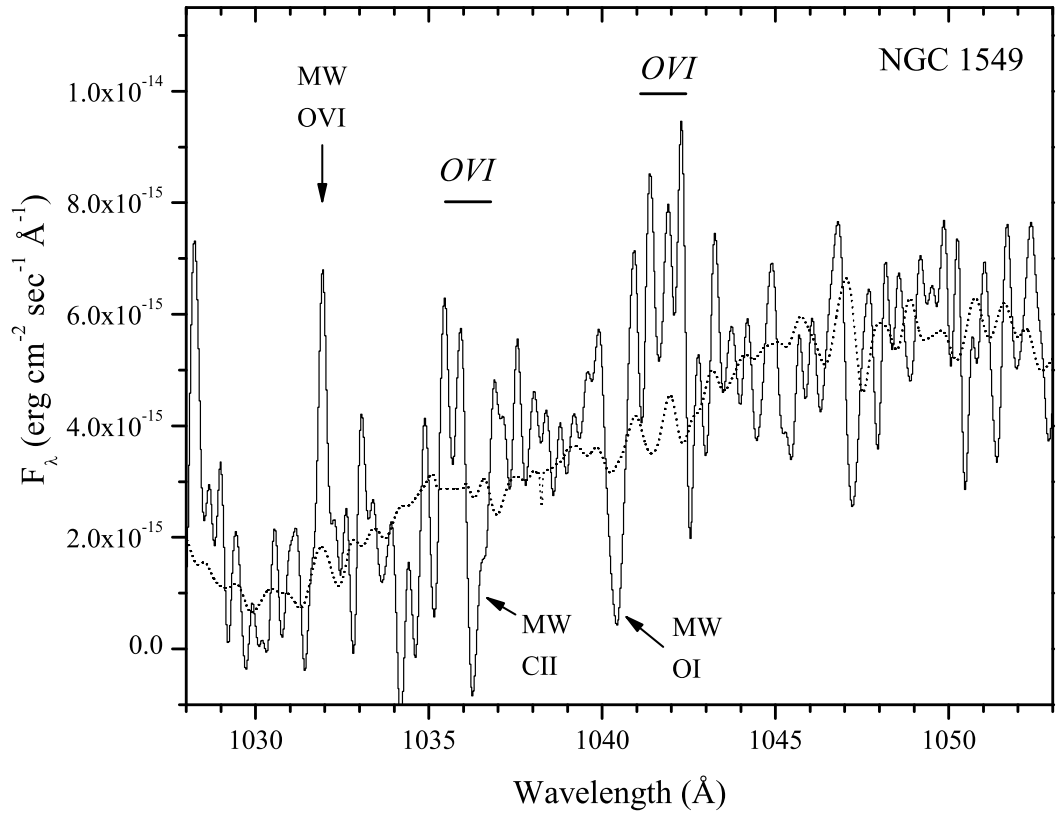}
\caption{OVI emission is detected in NGC 1549, with Galactic atomic absorption (CII) 
reducing the strength of the strong OVI line.\label{fig6}}
\end{figure}

\begin{figure}
\plotone{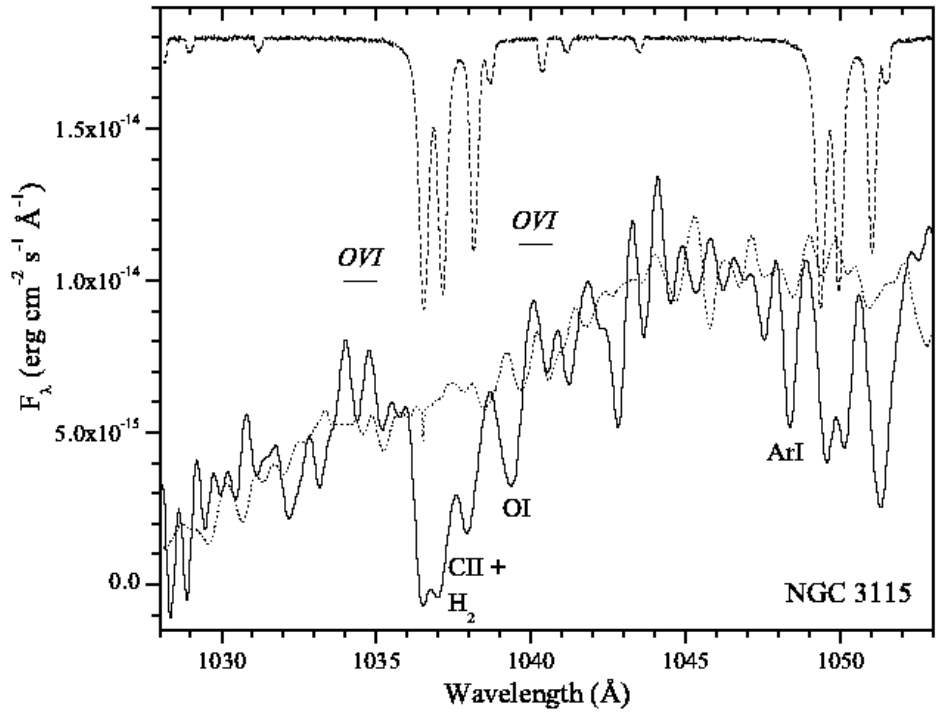}
\caption{The stellar continuum of NGC 3115 is well-defined as is the Galactic atomic and molecular
absorption, but the OVI emission falls below our 3${\sigma }$ threshold.\label{fig7}}
\end{figure}

\begin{figure}
\plotone{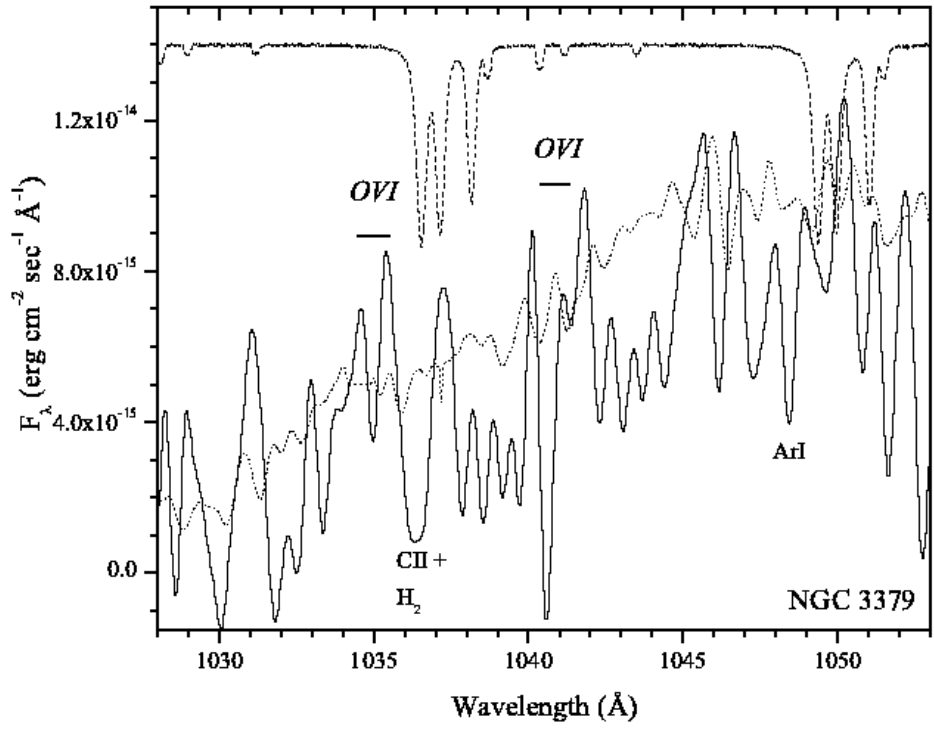}
\caption{The UV spectrum of NGC 3379 does not show any OVI emission; Galactic atomic
absorption is seen for one line.  The deep absorption at 1040.5 \AA \ is of 
unknown origin.\label{fig8}}
\end{figure}

\begin{figure}
\plotone{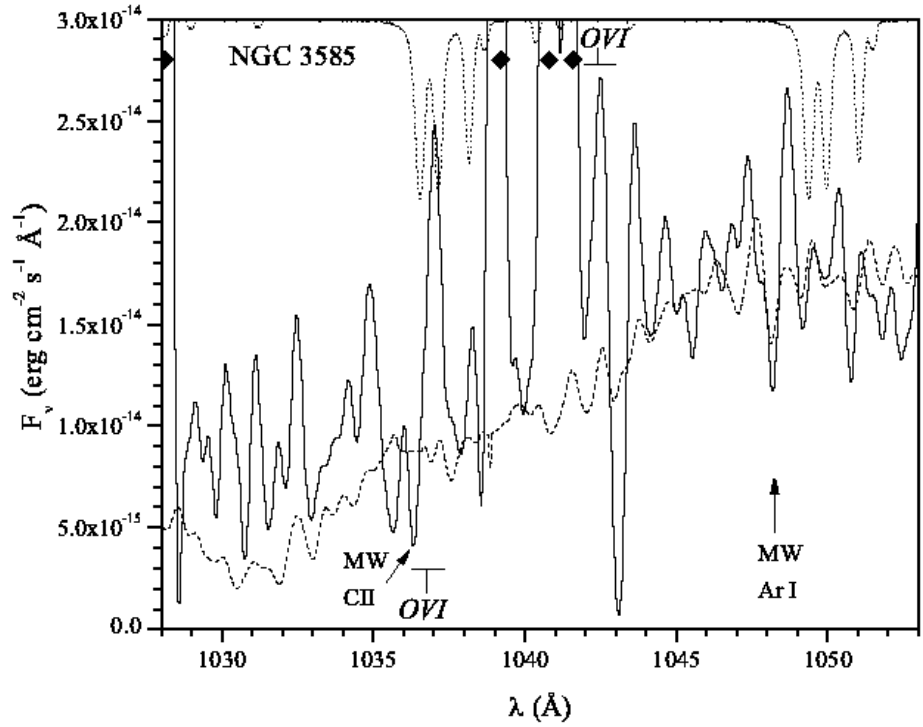}
\caption{Only observations during the day were useful and there was a very large
amount of scattered light and airglow lines (denoted by black diamonds), leading to a poor quality spectrum.  None
of the strong Galactic absorption features are detected, such as from CII.
No OVI emission from NGC 3585 is detected at a statistically significant level.
\label{fig9}}
\end{figure}

\begin{figure}
\plotone{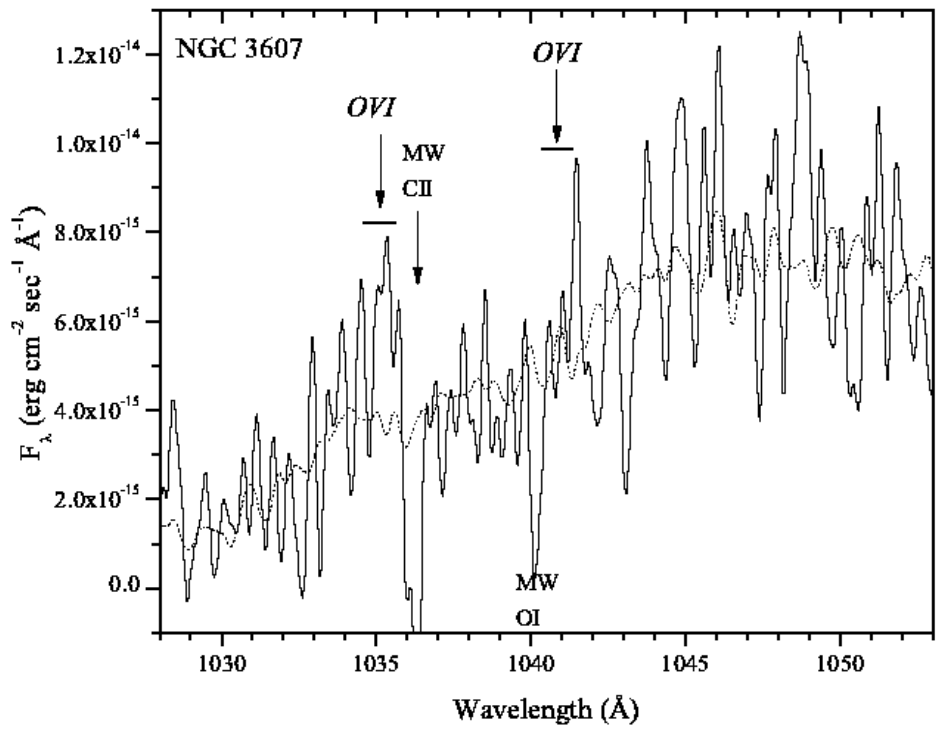}
\caption{For NGC 3607, the OVI line is detected at above the 3${\sigma} $ threshold.  
The usual strong Galactic absorption lines are present and the stellar continuum
closely follows that of NGC 1399 (dotted line).\label{fig10}}
\end{figure}

\begin{figure}
\plotone{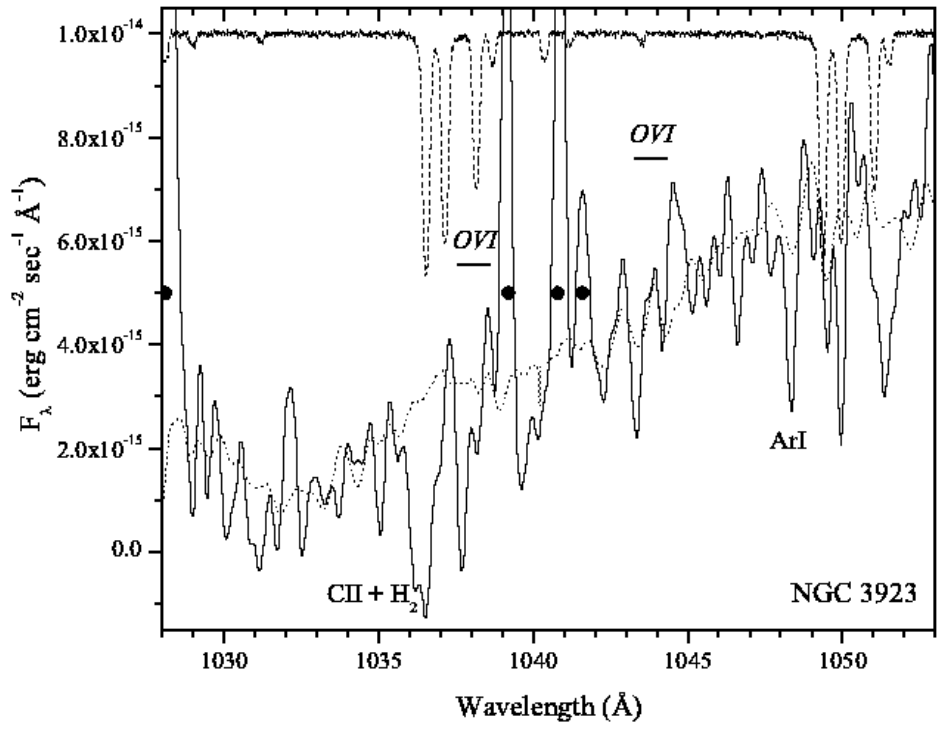}
\caption{Airglow lines were present in all spectra of NGC 3923 (solid 
circles).  No OVI emission is present in the stellar continuum that is
similar to NGC 1399 (dotted line), while Galactic atomic and molecular
lines are detected.\label{fig11}}
\end{figure}

\begin{figure}
\plotone{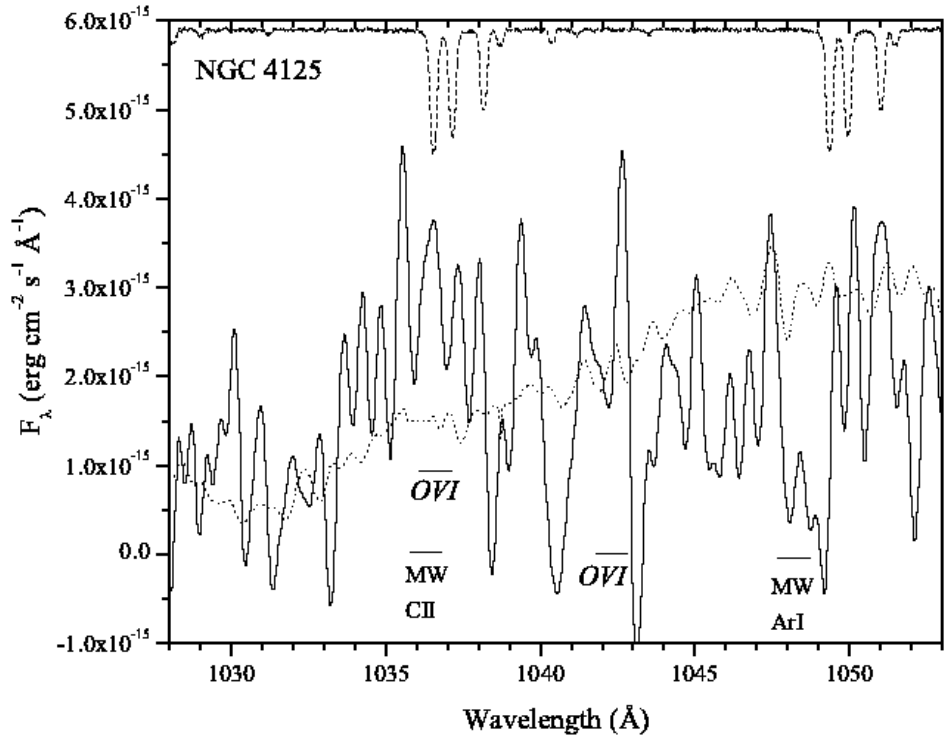}
\caption{The spectrum of NGC 4125 is hardly visible, and none of the strongest Galactic
absorption features are detected (e.g., CII), nor could a stellar continuum be
fit to the data at an acceptable level.  No statistically significant OVI
emission features are present.\label{fig12}}
\end{figure}

\begin{figure}
\plotone{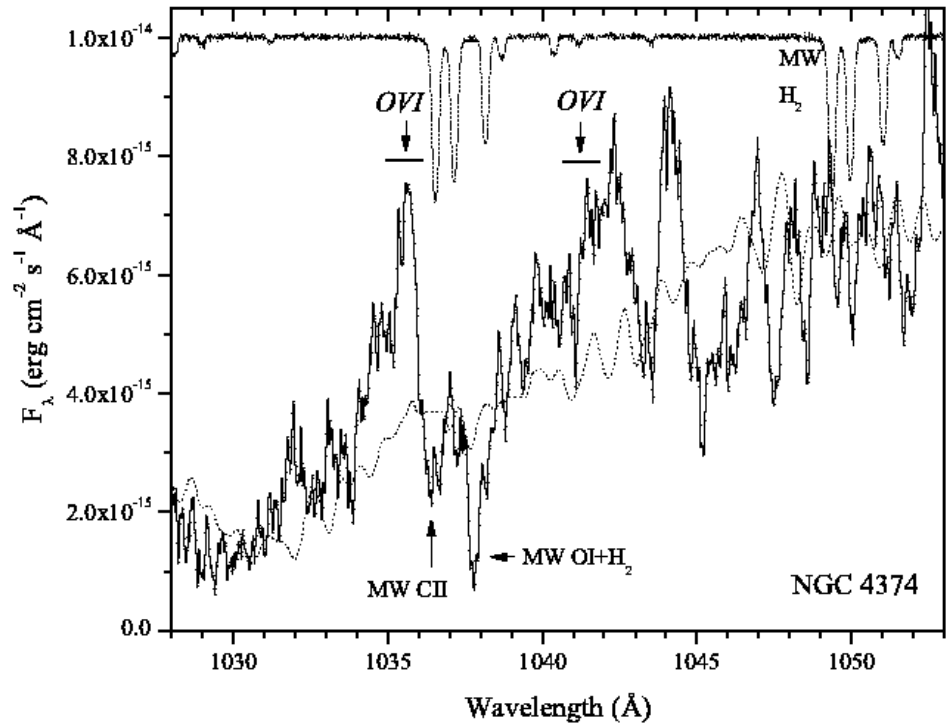}
\caption{Emission from OVI is detected in NGC 4374, although some of the
emission from the strong line may be absorbed by the Galactic CII line (an 
possibly H$_2$ as well).  The weak OVI line appears to be present but 
falls near an instrumental feature and may be distorted.\label{fig13}}
\end{figure}

\begin{figure}
\plotone{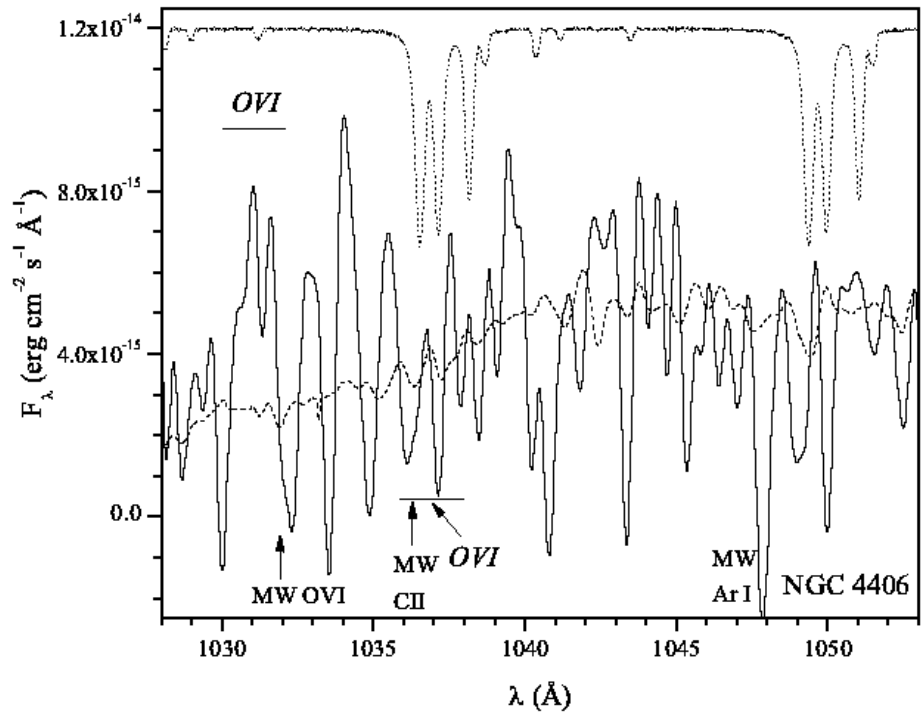}
\caption{A possible OVI emission line may be present in this spectrum of NGC 4406,
where the stellar continuum is so weak that the usual Galactic absorption 
features are not evident (CII).\label{fig14}}
\end{figure}

\begin{figure}
\plotone{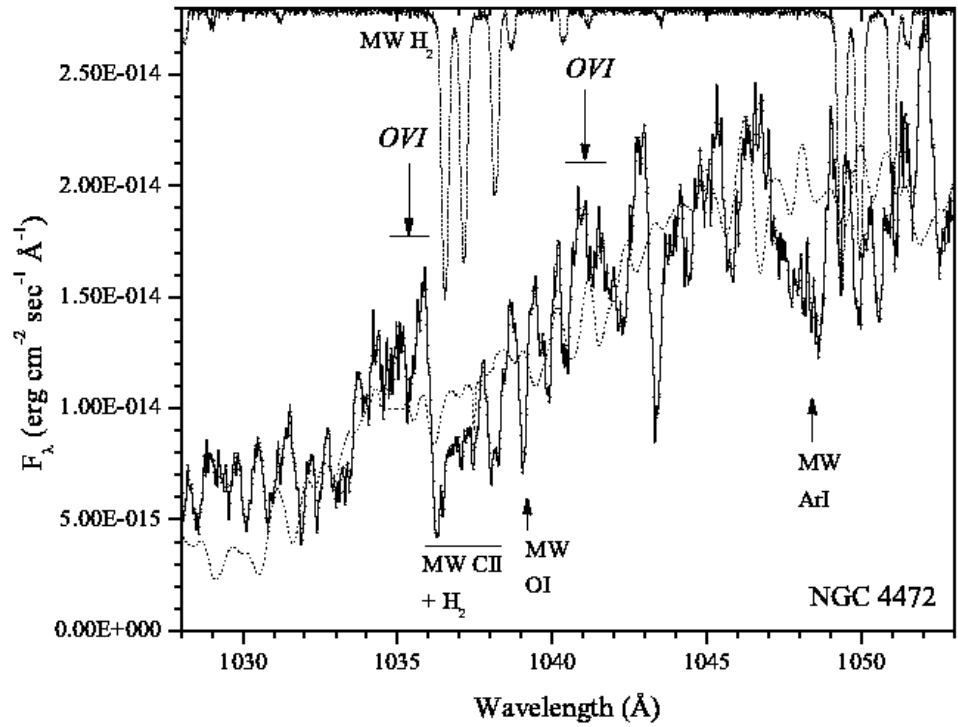}
\caption{Possible OVI emission from both lines is present near the 3${\sigma }$
threshold for NGC 4472.  Absorption by the Galactic CII line may have absorbed some of 
the red side of the strong OVI line.\label{fig15}}
\end{figure}
\clearpage

\begin{figure}
\plotone{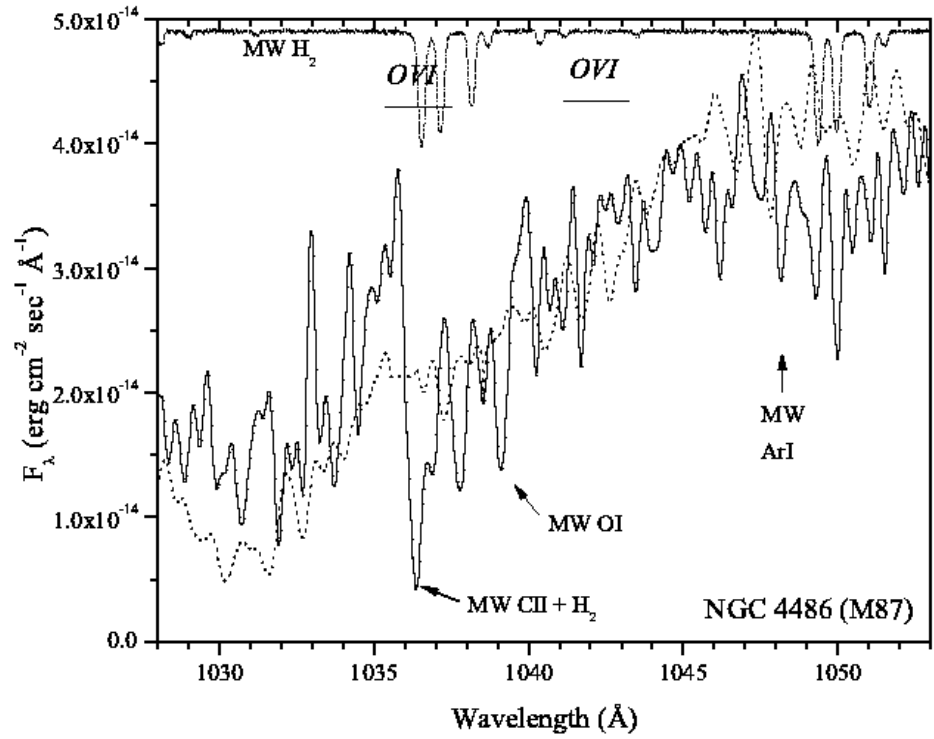}
\caption{Emission from the stronger OVI may be present in M87, although the red
side of the line would have been absorbed by Galactic CII and H$_2$ gas.\label{fig13}}
\end{figure}

\begin{figure}
\plotone{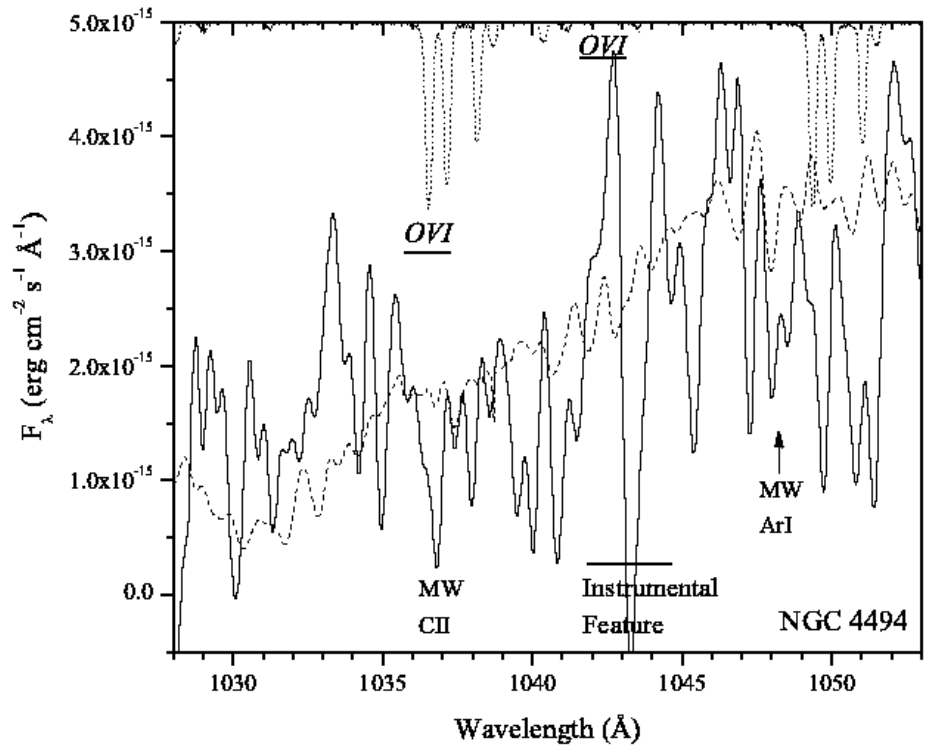}
\caption{The stellar continuum is weak in NGC 4494 and there is no evidence for 
the strong OVI line; the weaker OVI line falls on an instrumental feature.\label{fig17}}
\end{figure}

\begin{figure}
\plotone{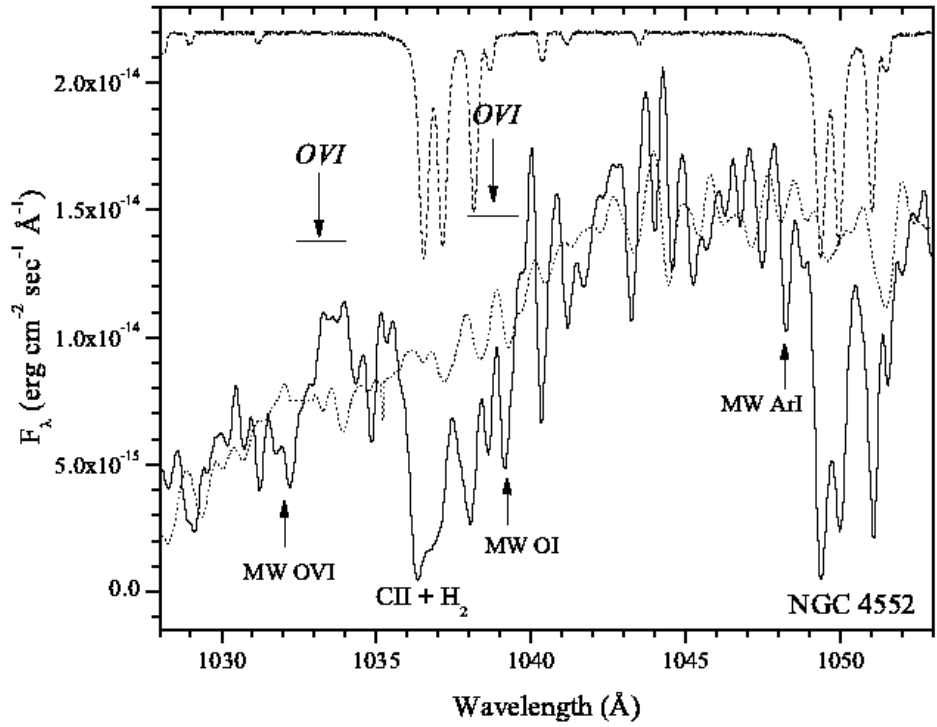}
\caption{The strong OVI emission is detected but the weaker OVI line 
would be absorbed by Galactic OI plus H$_2$ gas, which is plentiful 
along this sight line.\label{fig18}}
\end{figure}

\begin{figure}
\plotone{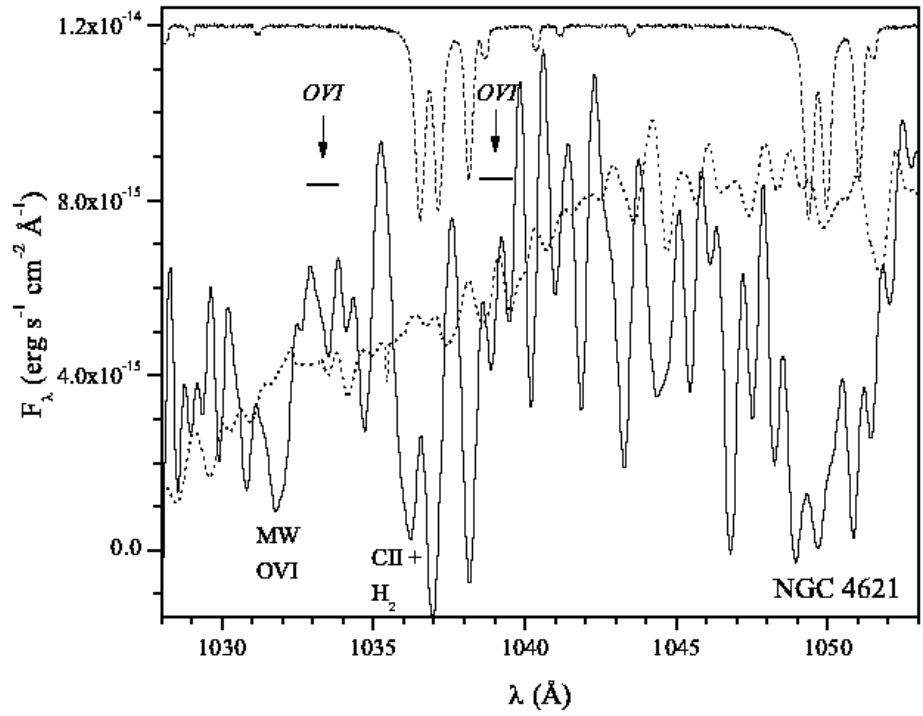}
\caption{The OVI emission is below the detection threshold, where the only
clear features are due to Galactic H$_2$ absorption.\label{fig19}}
\end{figure}

\begin{figure}
\plotone{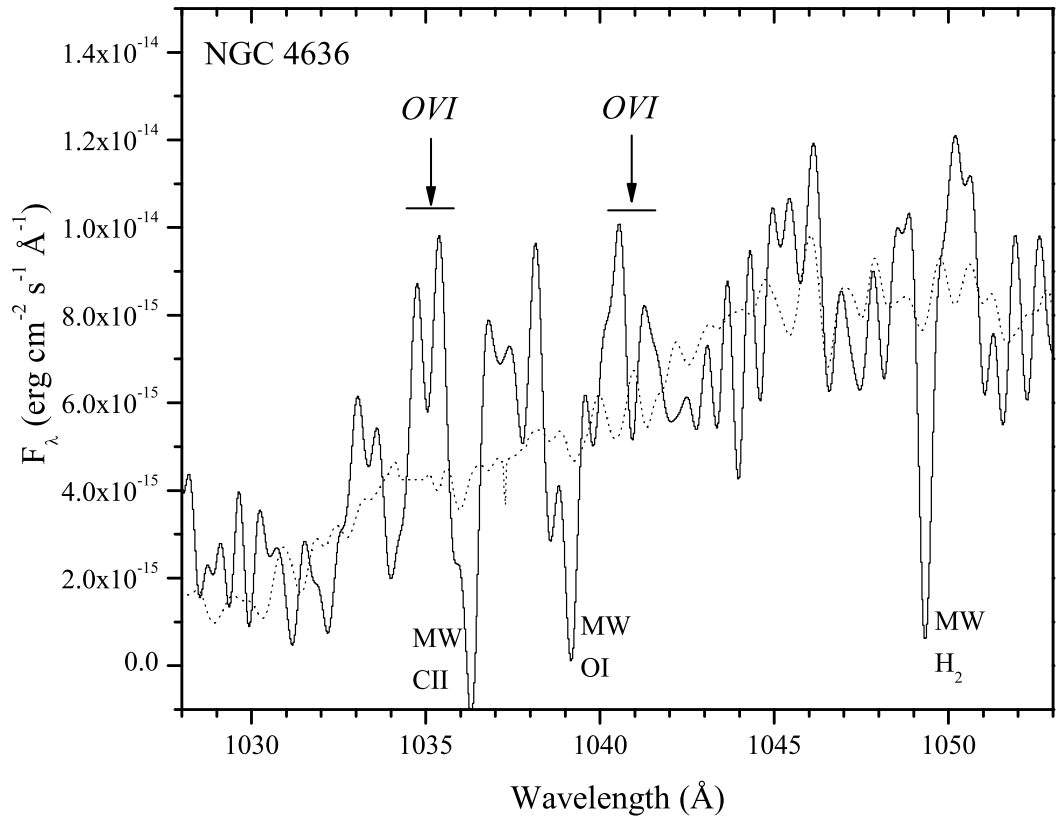}
\caption{The strongest Galactic absorption features are detected in
NGC 4636, as well as the OVI emission.\label{fig20}}
\end{figure}

\begin{figure}
\plotone{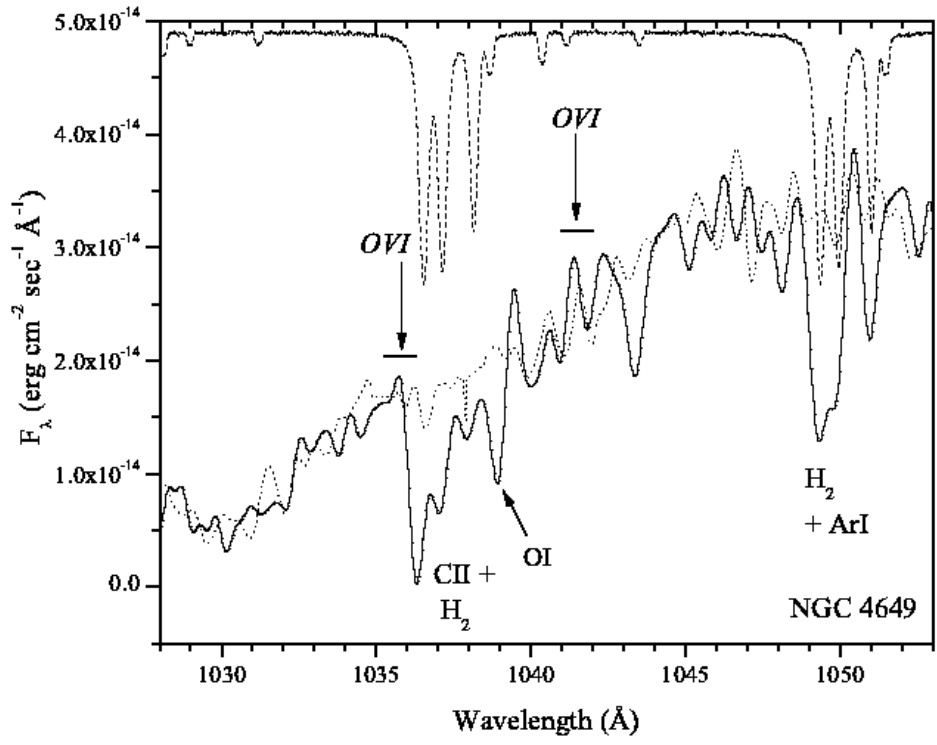}
\caption{No OVI emission is seen in this well-defined stellar continuum, 
which contains the usual Galactic absorption features.\label{fig21}}
\end{figure}
\clearpage

\begin{figure}
\plotone{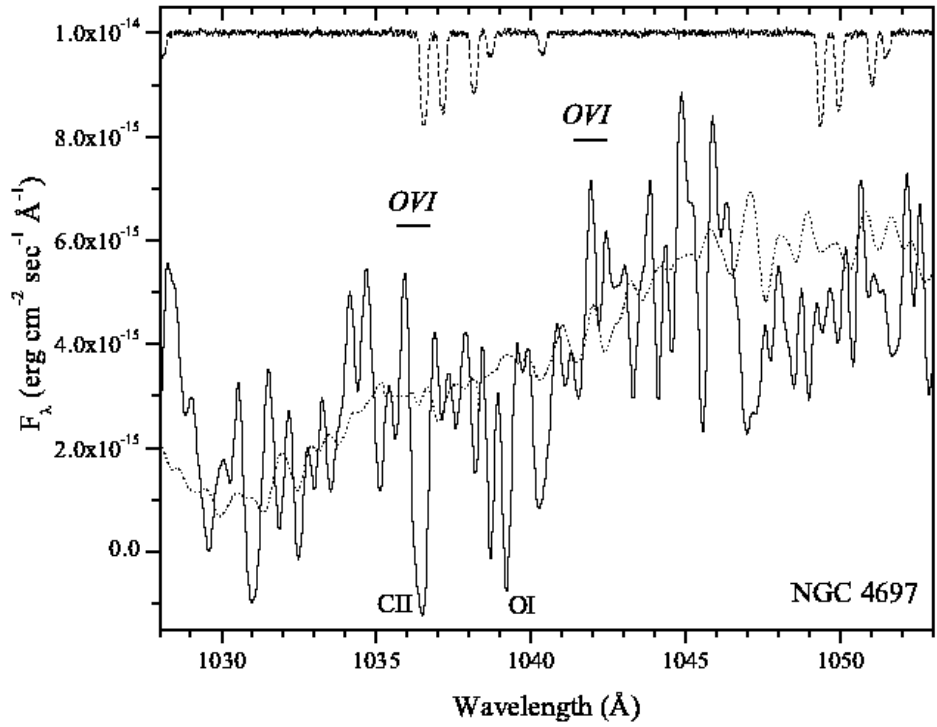}
\caption{No significant OVI emission is detected from this spectum, which 
only shows a stellar continuum plus Galactic atomic absorption lines.\label{fig22}}
\end{figure}

\begin{figure}
\plotone{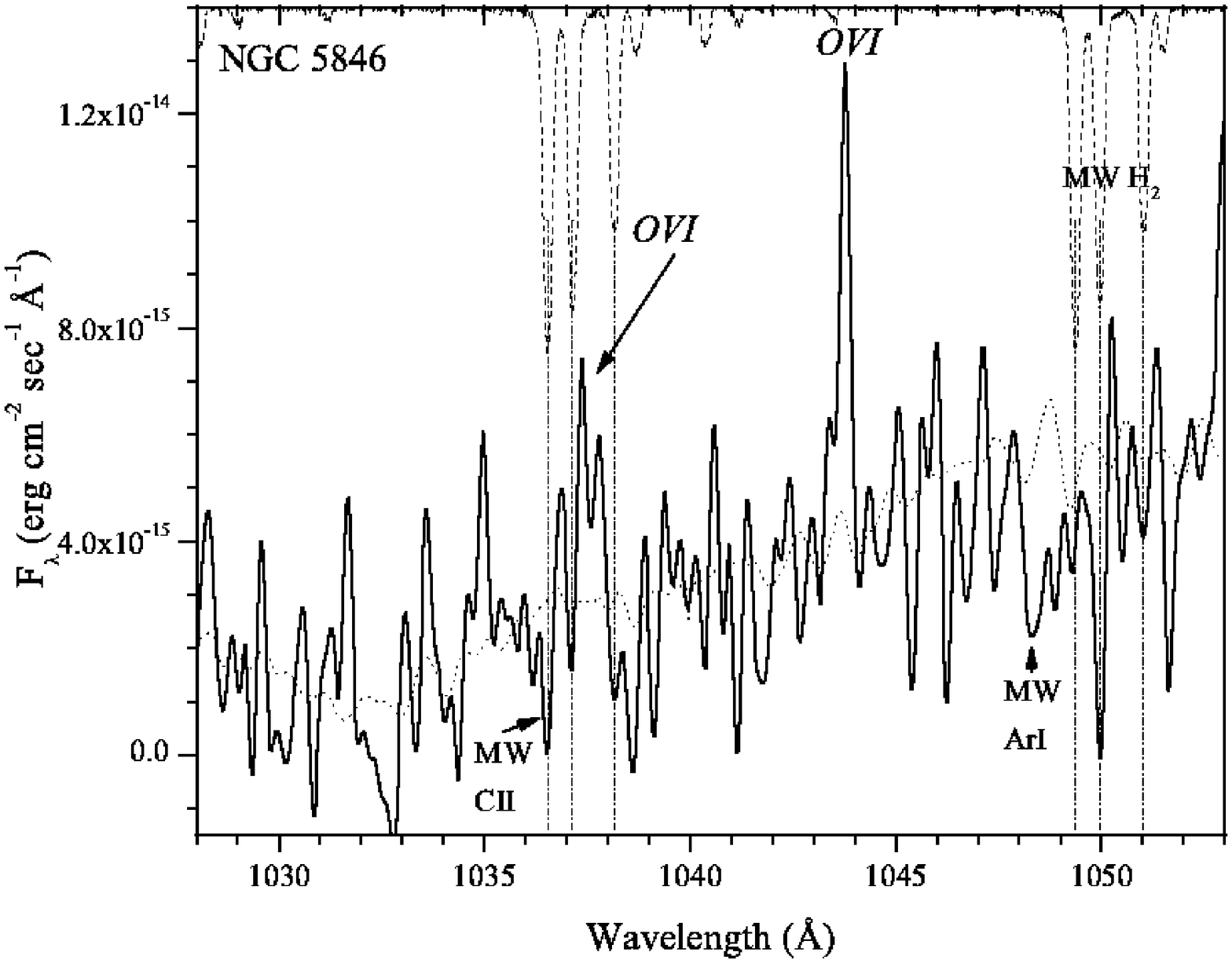}
\caption{The strong OVI line from NGC 5846 is partly absorbed by the Galactic
CII plus H$_2$ lines.\label{fig23}}
\end{figure}

\begin{figure}
\plotone{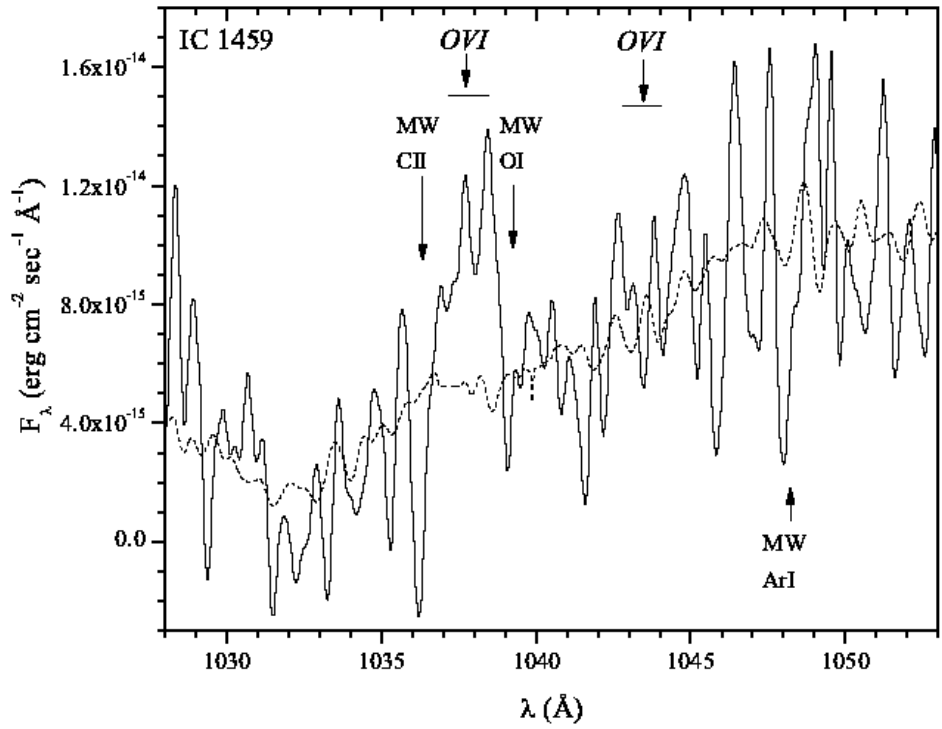}
\caption{The strong OVI line, detected in this spectrum of IC 1459, 
is probably partially absorbed by Galactic CII plus OI.  Galactic 
molecular absorption is not detected.\label{fig24}}
\end{figure}

\begin{figure}
\plotone{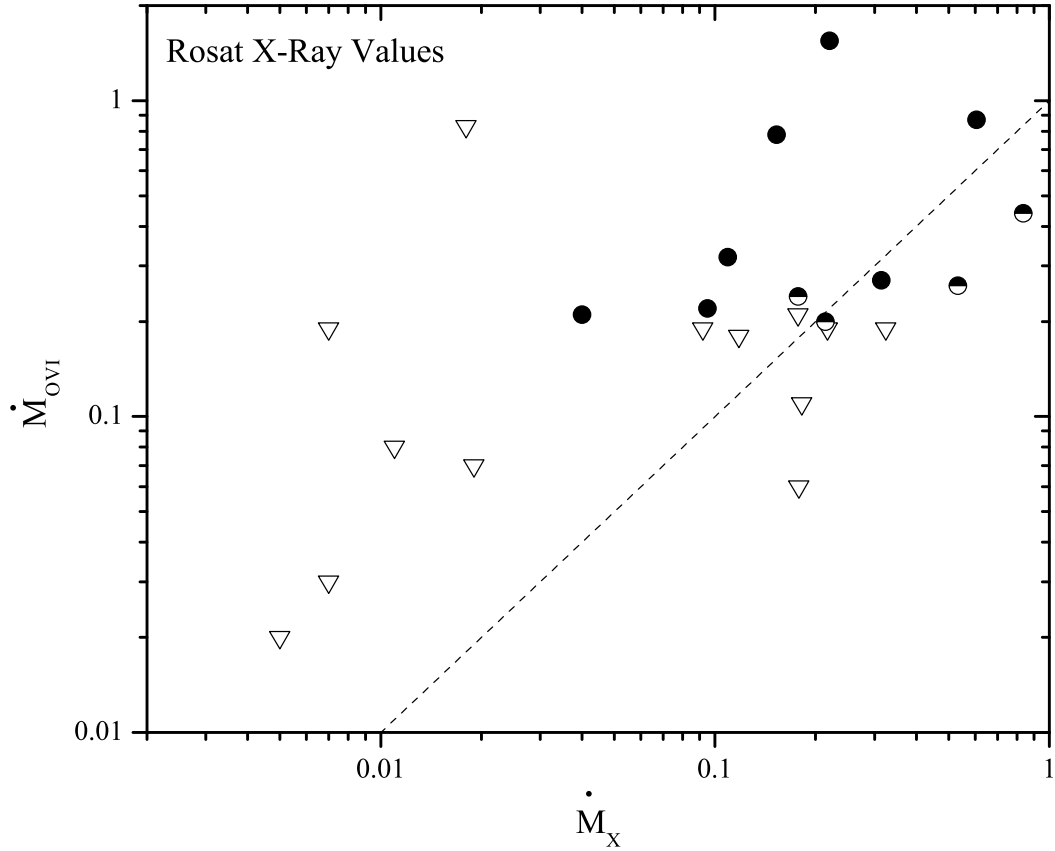}
\caption{The cooling rate determined from the OVI data (in M$_{\odot} $ yr$^{-1}$) 
vs the cooling rate from the {\it{ROSAT}\/} data.  The upper limits for the
OVI data are given by open triangles, the possible detections by half-filled
circles, and the detections by filled circles.  The {\it{ROSAT}\/} data
contain the emission from the gas as well as stellar sources.  The dashed
line shows ${\dot{M}}_{OVI}$ = ${\dot{M}}_X$, and is not a fit to the data.\label{fig25}}
\end{figure}

\begin{figure}
\plotone{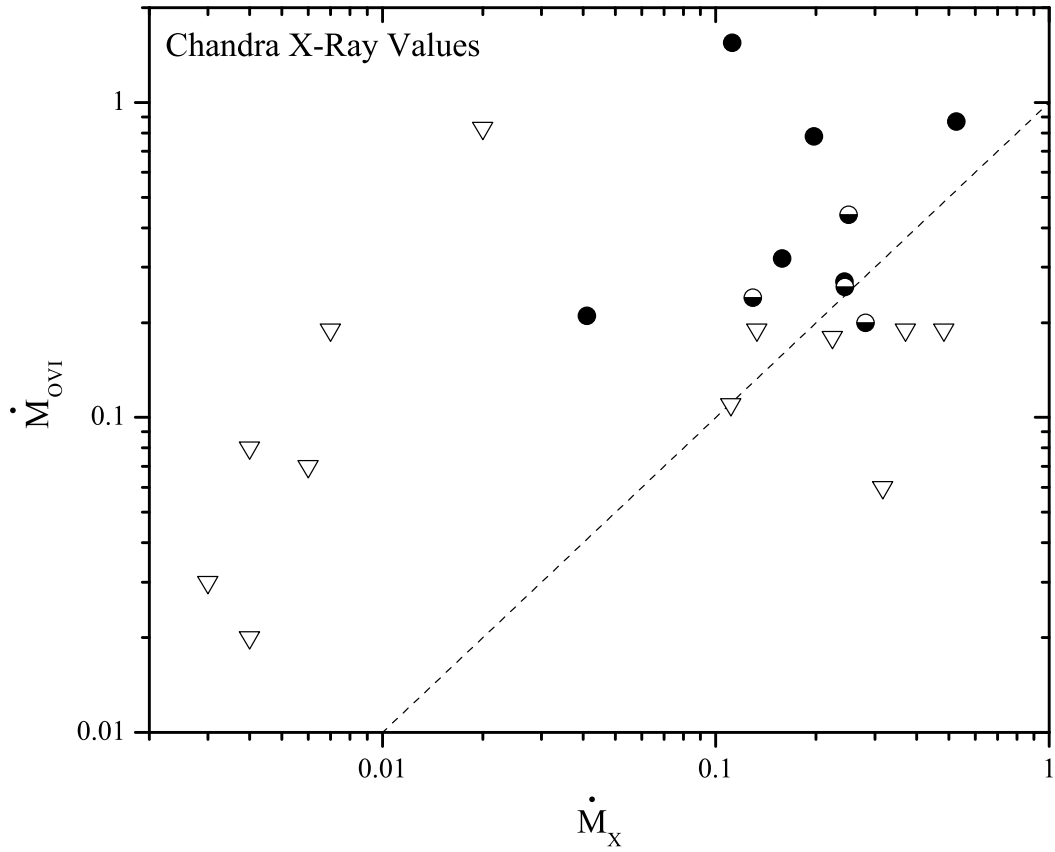}
\caption{Similar to the above figure, but where the X-ray cooling rates
are determined from the {\it Chandra\/} data, in which stellar emission 
has been removed.\label{fig26}}
\end{figure}

\end{document}